\documentclass{mn2e}
\usepackage{graphicx}

\begin{document}
\title{On the detection of pre-low-mass X-ray binaries}
\author[B. Willems and U. Kolb]
{B. Willems\thanks{E-mail: B.Willems@open.ac.uk, U.C.Kolb@open.ac.uk}
and U. Kolb$^\star$ \\
Department of Physics and Astronomy, The Open University,
Walton Hall, Milton Keynes, MK7 6AA, UK}

\date{Accepted ... Received ...; in original form ...}

\pagerange{\pageref{firstpage}--\pageref{lastpage}} \pubyear{2003}

\maketitle

\label{firstpage}

\begin{abstract}
We explore the population of candidate pre-low-mass X-ray binaries in
which a neutron star accretes mass from the wind of a low-mass
companion (mass $\le 2\,M_\odot$) in the framework of a binary
population synthesis study. The simulated accretion-luminosity
distribution shows a primary peak close to $10^{31}$ erg/s and a
secondary peak near $10^{28}$ erg/s. The relative contribution of the
two peaks depends primarily on the magnitude of the kick velocity
imparted to the neutron star at birth. The secondary peak is
negligible for average kick velocities larger than $\sim 200$ km/s,
but becomes dominant for average kick velocities smaller than $\sim
50$ km/s. Regardless of the relative contributions of the two peaks,
our calculations suggest that pre-low-mass X-ray binaries may provide
a non-negligible contribution to the population of discrete
low-luminosity X-ray sources in the Galaxy.
\end{abstract}


\section{Introduction}

Ongoing deep X-ray surveys of selected Galactic regions have sparked a
renewed interest in low-luminosity Galactic X-ray sources such as
wind-accreting neutron stars.  The XMM--Newton Galactic Plane Survey
(e.g.\ Warwick 2002), the XMM-Newton Serendipitous Survey (Watson
2001; Motch et al.\ 2002; Motch, Herent \& Guillout 2003; Watson et
al. 2003) and the
serendipitous Chandra Multi-wavelength Plane Survey (Grindlay et
al. 2003) will all probe the faint Galactic X-ray point source
population.

The discovery by Chandra of more than 500 previously undetected point
sources with luminosities above $\simeq 10^{35}$ erg/s in a $2.0^\circ
\times 0.8^\circ$ field near the Galactic centre (Wang, Gotthelf \&
Lang 2002) dramatically illustrates the potential of such surveys and
has already prompted complementing theoretical studies.  Pfahl,
Rappaport \& Podsiadlowski (2002) considered the population of neutron
stars that accrete from the wind of an intermediate- or high-mass star
(mass $\ge 3 \, M_\odot$). They found that these wind-driven
intermediate-mass and high-mass X-ray binaries may indeed account for
a dominant fraction of the point source population found by Wang et
al. (2002).

Independently, Bleach (2002) investigated the possibility of detecting
detached close binaries consisting of a neutron star and a red dwarf,
which may be the progenitors of low-mass X-ray binaries (LMXBs). The
proposed detection mechanism consists of looking for red dwarfs with
X-ray luminosities larger than those expected from intrinsic X-ray
emissions from the star's corona. The excess could then possibly be
ascribed to the emission of X-rays by a neutron star companion
accreting mass from the wind of the red dwarf.

Our aim in this paper is to elaborate on the idea by Bleach (2002) and
examine the detection probability of pre-LMXBs, i.e.\ detached
neutron-star systems with low-mass companions (mass $\le 2\, M_\odot$),
in the context of a binary population synthesis study.

\section{pre-LMXBs and LMXBs}

In close binary systems of stars, mass can be transferred from one
component to the other when one of the stars loses mass in a stellar
wind or when one of the stars fills its Roche lobe. If the companion
star is a compact object which is able to accrete some of the
transferred mass, the liberation of gravitational potential energy
gives rise to an accretion luminosity
\begin{equation}
L_{\rm acc} = {{G M \dot{M}_{\rm acc}} \over R},
  \label{Lacc}
\end{equation}
where $G$ is the Newtonian constant of gravitation, $M$ and $R$ are
the mass and the radius of the accreting compact object, and
$\dot{M}_{\rm acc}$ is the mass-accretion rate. We assume 
that the accretion process yields an X-ray luminosity $L_{\rm X}
\approx L_{\rm acc}$.

In the particular case of a wind-accreting neutron star, the spin
evolution of the neutron star and the possible inhibition of accretion
by strong centrifugal forces may play an important role in the study
of the accretion process (e.g. Pringle \& Rees 1972; Illarionov \&
Sunyaev 1975; Stella, White, Rosner 1986; Urpin, Geppert \& Konenkov
1998). The treatment of these inhibiting effects is however beyond the
scope of this initial investigation. Instead, we assume that at the
onset of mass transfer via the stellar wind, the neutron star's
radiation pressure and rotation rate is low enough for mass accretion
to be possible.

The evolution of binaries containing a neutron star is driven by the
nuclear evolution of the companion star or by the loss of orbital
angular momentum via magnetic braking and/or gravitational
radiation. When the combined effect of these processes causes the
companion to become larger than its Roche-lobe, an X-ray binary may be
formed. The further evolution of the binary depends critically on the
orbital period and the mass of the donor star at the onset of
Roche-lobe overflow. The formation and evolution of X-ray binaries
with low-mass donor stars have been discussed extensively in the past
by, e.g., Verbunt (1993); Verbunt \& van den Heuvel (1995); Iben \&
Tutukov (1995); Kalogera \& Webbink (1996, 1998); Kalogera (1998);
Kalogera, Kolb \& King (1998); and more recently by Podsiadlowski,
Rappaport \& Pfahl (2002). We therefore refer to these studies and
references therein for detailed discussions on the various formation
channels and here consider the population of X-ray binaries descending
from the different channels as a whole.

For the purpose of this investigation, we define a candidate pre-LMXB
as a detached binary in which a neutron star accretes mass from the
stellar wind of its companion. We assume the latter to be on the main
sequence, in the Hertzsprung gap, or on the giant branch, and to have
a mass smaller than or equal to $2\,M_\odot$. The luminosity
associated with the accretion process is required to be larger than
$10^{-6}\,L_\odot$ or ten times the mass donor's coronal X-ray
luminosity, whichever is larger. We note that our definition of
pre-LMXBS differs from that of Bleach (2002) in that we consider stars
with masses up to $2\,M_\odot$ and binaries that are expected to
become semi-detached due to the nuclear evolution of the secondary as
well as those that are expected to become semi-detached due to angular
momentum losses via magnetic braking and/or gravitational radiation.

Similarly, we define a LMXB as a semi-detached system in which a
neutron star accretes mass from a Roche-lobe filling main-sequence,
Hertzsprung-gap, or giant-branch star with a mass smaller than or
equal to $2\,M_\odot$. In order to be classified as an X-ray binary,
we furthermore require the mass-accretion rate to be high enough to
produce an X-ray luminosity in excess of $10^{-1}\,L_\odot$.

For the remainder of the paper, we denote the mass and the radius of
the neutron star by $M_{\rm NS}$ and $R_{\rm NS}$, and the mass and
the radius of the donor star by $M_{\rm d}$ and $R_{\rm d}$,
respectively.

\section{Assumptions and computational technique}

We explore the possibility of detecting and identifying LMXBs prior to
the onset of Roche-lobe overflow from the neutron star's companion by
means of the BiSEPS binary population synthesis code described by
Willems \& Kolb (2002). The code is based on the analytic
approximation formulae for the evolution of single stars derived by
Hurley, Pols \& Tout (2000) and follows the main lines of the binary
evolution algorithm described by Hurley, Tout \& Pols (2002). The
binary orbits are assumed to be circular and the stellar rotation
rates are kept synchronised with the orbital motion at all times. We
furthermore restrict ourselves to Population~I stellar compositions.

\begin{figure}
\resizebox{8.0cm}{!}{\includegraphics{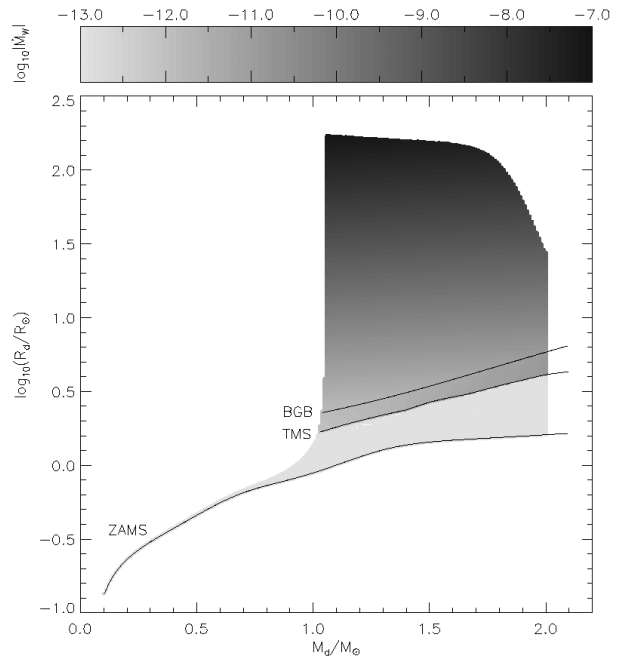}}
\caption{The logarithm of the wind mass-loss rates (in $M_\odot\,{\rm
  yr}^{-1}$) used in our calculations, as a function of the stellar
  mass and radius. The lines labelled ZAMS, TMS, and BGB represent the
  radii of the stars at the zero-age main sequence, the terminal main
  sequence, and the base of the giant branch, respectively. }
\label{wind}
\end{figure}

For the purpose of this investigation, the mass-loss rates from
stellar winds given by Hurley et al. (2000), which are limited to
evolved stars and massive main-sequence stars, must be supplemented by
a prescription for the winds from low-mass main-sequence
stars. Although these winds are not fully understood yet, it is
generally accepted that they depend sensitively on both the mass and
the age of the donor star (e.g. Wood et al. 2002, and references
therein). Young solar-type stars and M dwarfs, for instance, are
thought to have mass-loss rates which may be as high as
$10^{-12}\,M_\odot$ per year, while older G dwarfs such as the Sun
have mass-loss rates of the order of $10^{-14}\,M_\odot$ per year
(Hartmann 1985, Lim \& White 1996, van den Oord \& Doyle 1997,
Wargelin \& Drake 2001, Sackmann \& Boothroyd 2003).  In view of these
still existing uncertainties, we here simply adopt a constant
mass-loss rate of $10^{-13}\,M_\odot$ per year for low-mass
main-sequence stars (see also Bleach 2002). The resulting mass-loss
rates are shown in Fig.~\ref{wind} for donor stars with masses up to
$2\,M_\odot$ and evolutionary stages from the zero-age main sequence
up to the tip of the giant branch. The lines labelled ZAMS, TMS, and
BGB represent the radii of the stars at the zero-age main sequence,
the terminal main sequence, and the base of the giant branch,
respectively. The maximum evolutionary age considered for the
calculation of the mass-loss rates is 10\,Gyr, so that only stars with
masses larger than $1\,M_\odot$ significantly evolve away from the
zero-age main sequence.

We estimate the mean mass-accretion rate onto a star orbiting in the
stellar wind of its companion by means of a standard
Bondi-Hoyle-Lyttleton formalism (Hoyle \& Lyttleton 1941, Bondi \&
Hoyle 1944). The accretion rate for binaries with circular orbits
then depends on the orbital separation $a$, the binary mass ratio $q$,
the donor star's radius $R_{\rm d}$, the wind mass-loss rate
$\dot{M}_{\rm d}$, and the wind velocity $v_{\rm w}$. We parametrise the
latter as a fraction $\beta_{\rm w}$ of the escape velocity $v_{\rm
e}= \left( 2GM_{\rm d}/R_{\rm d} \right)^{1/2}$ at the surface of the
mass-losing star (see, e.g., Hurley et al. 2002). The mean wind
mass-accretion rate can then be cast in the form
\begin{equation}
\dot{M}_{\rm acc} = {3 \over {16}} \left( {R_{\rm d} \over a} \right)^2
  {q^2 \over \beta_{\rm w}^4} \left( 1 + {{1 + q}
  \over {2\, \beta_{\rm w}^2}} {R_{\rm d} \over a}
  \right)^{-3/2} \left| \dot{M}_{\rm d} \right|. \label{Macc2}
\end{equation}
In the particular case of a wind-accreting neutron star the mass ratio
is given by $q=M_{\rm NS}/M_{\rm d}$.

The detection mechanism for pre-LMXBs proposed by Bleach (2002)
furthermore requires a prescription for the intrinsic X-ray luminosity
of the neutron star's companion. Since this luminosity is thought to have
its origin in dynamo processes generated by the interaction between
convection and rotation, it can be expected to be correlated
with the star's rotational angular velocity (e.g. Belvedere, Chiuderi
\& Paterno 1982; Noyes et al. 1984). The existence of such a
correlation was explored extensively from an observational point of
view by, e.g., Pallavicini et al. (1981); Marilli \& Catalano (1984);
Vilhu (1984); Doyle (1987); Fleming, Gioia \& Maccacaro (1989);
Hempelmann et al. (1995); Randich et al. 1996; Singh et al. (1999);
Gondoin (1999). The numerous investigations however show a clear lack
of consensus on whether or not the correlation exists and, if it
exists, what the precise relationship is between the rotational
angular velocity and the coronal X-ray luminosity. This deficiency may
be related to the use of different stellar samples and observational
selection criteria (Hempelmann et al. 1995, Gondoin 1999).

A point of lesser dispute seems to be the existence of an upper limit
on the intrinsic stellar X-ray luminosity, although its origin remains
to be unambiguously identified (Walter 1982, Byrne et al. 1984, Vilhu
\& Walter 1987, Jardine \& Unruh 1999, Singh et al. 1999). The upper
limit may be related to the finite size of the stellar surface in the
sense that the magnetic activity can only increase with increasing
rotational angular velocities until the whole stellar surface is
covered with magnetically active regions. Based on this idea,
Fleming et al. (1989) showed that the coronal saturation luminosity
$L_{\rm cor}$ could be related to the square of the stellar radius
$R_{\rm d}$ as
\begin{equation}
\log L_{\rm cor} = -2.9 + 2\, \log R_{\rm d}, \label{Lcor}
\end{equation}
where both $L_{\rm cor}$ and $R_{\rm d}$ are expressed in solar units
(see their Fig.~3).

Rather than face the uncertainties regarding the relation between
stellar rotation and coronal X-ray activity, we here opt to use a
prescription for the upper limit on the coronal X-ray luminosity.  We
use the mass of the convective envelope, $M_{\rm conv}$, to
distinguish between coronally active and non-active stars and,
somewhat arbitrarily, set the limit for coronal X-ray activity at
$M_{\rm conv} = 0.01\,M_{\rm d}$. Stars with smaller convective
envelopes are assumed to have negligible coronal X-ray luminosities,
while stars with larger convective envelopes are assumed to have
coronal X-ray luminosities given by Eq.~(\ref{Lcor}). The upper
limit may significantly overestimate the coronal X-ray luminosity of
giant-type stars with slow rotation rates and correspondingly weak
dynamos (Schr\"{o}der, H\"unsch \& Schmitt 1998; Gondoin 1999). The
slow rotation, however, only applies to giants in wide binaries where
tidal forces are negligible. Giants in closer binaries are subjected
to tidal torques which tend to synchronise the giant's rotation with
the orbital motion of the companion. A significant dynamo action may
therefore be sustained in these stars during much more evolved stages
of stellar evolution than for single stars or stars in wide
binaries. Dempsey et al. (1993, 1997) have shown that in the
particular cases of RS Canum Venaticorum (RS CVn) and BY Draconis (BY
Drac) binaries the 
coronal X-ray luminosity of the giant-type component may still reach
values up to $0.1\,L_\odot$. To be on the safe side, we therefore
still use the upper limit given by Eq.~(\ref{Lcor}) for stars that
have evolved beyond the main-sequence. Where relevant, we will discuss
the possible implications of this on our results. In any case, we
expect the consistent use of the upper limit for the coronal X-ray
luminosity to imply that our selection criteria for the identification
of pre-LMXBs are systematically too strict.

\section{Population synthesis}

We evolved a large number of binaries with initial component masses
ranging from $0.1$ to $60\,M_\odot$ and with initial orbital periods
between 0.1 to 10\,000 days. We used logarithmically spaced grids
consisting of 60 grid points for the initial stellar masses and 200
grid points for the initial orbital periods. The maximum evolutionary
age considered was 10\,Gyr. If a binary undergoes an asymmetric
supernova explosion leading to the birth of a neutron star, 200 random
kick velocities are generated from a Maxwellian velocity distribution
for the magnitude of the kick velocities. The evolution of the
binaries surviving the explosion is then followed until the imposed
age limit of 10\,Gyr.

The outcome of the binary evolution calculations depends on the
parameters adopted for the various evolutionary processes a binary may
be subjected to. We therefore vary the input parameters one at a time
and compare the resulting populations of pre-LMXBs with that of a
reference model. For the latter, hereafter referred to as model~A, we
adopt the same set of input parameters as in Willems \& Kolb (2002),
i.e. the common-envelope ejection efficiency parameter takes the value
$\alpha_{\rm CE}=1.0$, and the Maxwellian kick-velocity dispersion takes
the value $\sigma_{\rm kick}=190$ km/s. In addition, we here adopt a
standard value of 1.0 for the wind-velocity parameter $\beta_{\rm w}$.
The different parameters adopted in the other population synthesis
models are summarised in Table~\ref{models}. In the case of the
different supernova-kick models, the average kick velocities
$\overline{v}_{\rm kick}$ adopted in the Maxwellian velocity
distributions are listed in addition to the velocity dispersions
$\sigma_{\rm kick}$. We also note that the variations in the
wind-velocity parameter $\beta_{\rm w}$ considered in models MA1 and
MA2 are equivalent to considering variations in the wind mass-loss
rate $\dot{M}_{\rm d}$, but that the parameter $\beta_{\rm w}$ is a
more appropriate population synthesis parameter since it has a greater
impact on the determination of the mean mass-accretion rate. This is
readily illustrated by means of Eq.~(\ref{Macc2}) which shows that, to
a first approximation, an increase in $\beta_{\rm w}$ by a factor of
$2$ corresponds to a decrease in $\dot{M}_{\rm d}$ by a factor
of the order of $10$. Models COR1, COR2, and COR3 are introduced to
assess the sensitivity of our results to the prescription for the
coronal X-ray luminosity. In particular, the use of a constant value
for  $L_{\rm cor}$ in models COR1 and COR2 allows us to account for
the possible effects of overestimating the coronal X-ray luminosity of
giant-type stars by the use of Eq.~(\ref{Lcor}). 

\begin{table*}
\caption{Population synthesis model parameters. }
\label{models}
\begin{tabular}{lcccccc}
\hline
model & $\alpha_{\rm CE}$ & $\beta_{\rm w}$ & $\sigma_{\rm kick}$ & $\overline{v}_{\rm kick}$ & $L_{\rm cor}/L_\odot$ & $M_{\rm conv}/M_{\rm d}$ \\
\hline
A     & 1.0 & 1.0 & 190 km/s & 330 km/s & Eq. (\ref{Lcor}) & 0.01 \\
K0    & 1.0 & 1.0 & no kicks &     -    & Eq. (\ref{Lcor}) & 0.01 \\
KM25  & 1.0 & 1.0 & 25 km/s  &  43 km/s & Eq. (\ref{Lcor}) & 0.01 \\
KM50  & 1.0 & 1.0 & 50 km/s  &  87 km/s & Eq. (\ref{Lcor}) & 0.01 \\
KM75  & 1.0 & 1.0 & 75 km/s  & 130 km/s & Eq. (\ref{Lcor}) & 0.01 \\
KM100 & 1.0 & 1.0 & 100 km/s & 173 km/s & Eq. (\ref{Lcor}) & 0.01 \\
KM300 & 1.0 & 1.0 & 300 km/s & 520 km/s & Eq. (\ref{Lcor}) & 0.01 \\
KM400 & 1.0 & 1.0 & 400 km/s & 693 km/s & Eq. (\ref{Lcor}) & 0.01 \\
CE1   & 0.2 & 1.0 & 190 km/s & 330 km/s & Eq. (\ref{Lcor}) & 0.01 \\
CE2   & 5.0 & 1.0 & 190 km/s & 330 km/s & Eq. (\ref{Lcor}) & 0.01 \\
MA1   & 1.0 & 0.5 & 190 km/s & 330 km/s & Eq. (\ref{Lcor}) & 0.01 \\
MA2   & 1.0 & 2.0 & 190 km/s & 330 km/s & Eq. (\ref{Lcor}) & 0.01 \\
COR1  & 1.0 & 1.0 & 190 km/s & 330 km/s & $10^{-2}$ (constant) & 0.01 \\
COR2  & 1.0 & 1.0 & 190 km/s & 330 km/s & $10^{-4}$ (constant) & 0.01 \\
COR3  & 1.0 & 1.0 & 190 km/s & 330 km/s & Eq. (\ref{Lcor}) & 0.1 \\
\hline
\end{tabular}
\end{table*}

The contribution of a system to the population of pre-LMXBs
depends on the probability distribution functions of its initial
parameters and on the time the system spends as a member of the
population. We assume the initial mass $M_1$ of the primary to be
distributed according to the initial mass function
\begin{equation}
\renewcommand{\arraystretch}{1.4} \xi \left(M_1 \right) = \left\{
  \begin{array}{ll}
  0 & \hspace{0.3cm} M_1/M_\odot < 0.1, \\ 0.38415\, M_1^{-1} &
\hspace{0.3cm} 0.1 \le M_1/M_\odot < 0.75, \\ 0.23556\, M_1^{-2.7} &
\hspace{0.3cm} 0.75 \le M_1/M_\odot < \infty,
  \end{array}
\right. \label{imf}
\end{equation}
the initial mass ratio $q=M_2/M_1$ according to
\begin{equation}
\renewcommand{\arraystretch}{1.4} n(q) = \left\{
  \begin{array}{ll}
  1 & \hspace{0.3cm} 0 < q \le 1, \\ 0 & \hspace{0.3cm} q > 1,
  \end{array}
\right. \label{imrd}
\end{equation}
and the initial orbital separation $a$ according to
\begin{equation}
\renewcommand{\arraystretch}{1.4} \chi (a) = \left\{
  \begin{array}{ll}
  0 & a/R_\odot < 3 \mbox{ or } a/R_\odot > 10^4, \\ 0.12328\, a^{-1}
& 3 \le a/R_\odot \le 10^4
  \end{array}
\right. \label{iosd}
\end{equation}
(see Willems \& Kolb 2002, and references therein).

Each time a binary from our initial grid of parameters evolves into a
pre-LMXB, its weighted contribution is added to the probability
density function (PDF) describing the distribution of newborn
pre-LMXBs in the $(t,M_{\rm d},P_{\rm orb},L_{\rm acc})$-space at the
time $t=t_b$ of its formation. The finite lifetime of the pre-LMXBs is
taken into account by subtracting the same weighted contribution from
the distribution function at the time $t=t_d$ when the system ceases
to be a pre-LMXB. In doing so, we implicitly assume that the stellar
and orbital parameters of a pre-LMXB do not change significantly once
it is formed. We will discuss the circumstances under which this
assumption breaks down and the possible implications thereof in the
final section of the paper. For now, we remark that this assumption
does not affect the determinations of the birthrates and the total
number of systems obtained after integration over all system parameters.

The resulting distribution function for the population of pre-LMXBs is
subsequently convolved with a constant star-formation rate and
normalised so that the integral over all systems found is equal to
one.

\section{Donor star masses and orbital periods}

\begin{figure*}
\resizebox{5.83cm}{!}{\includegraphics{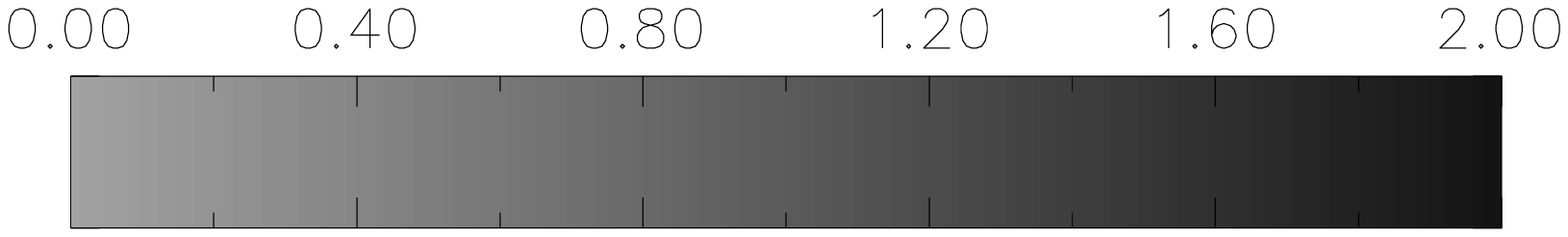}} \\
\resizebox{5.83cm}{!}{\includegraphics{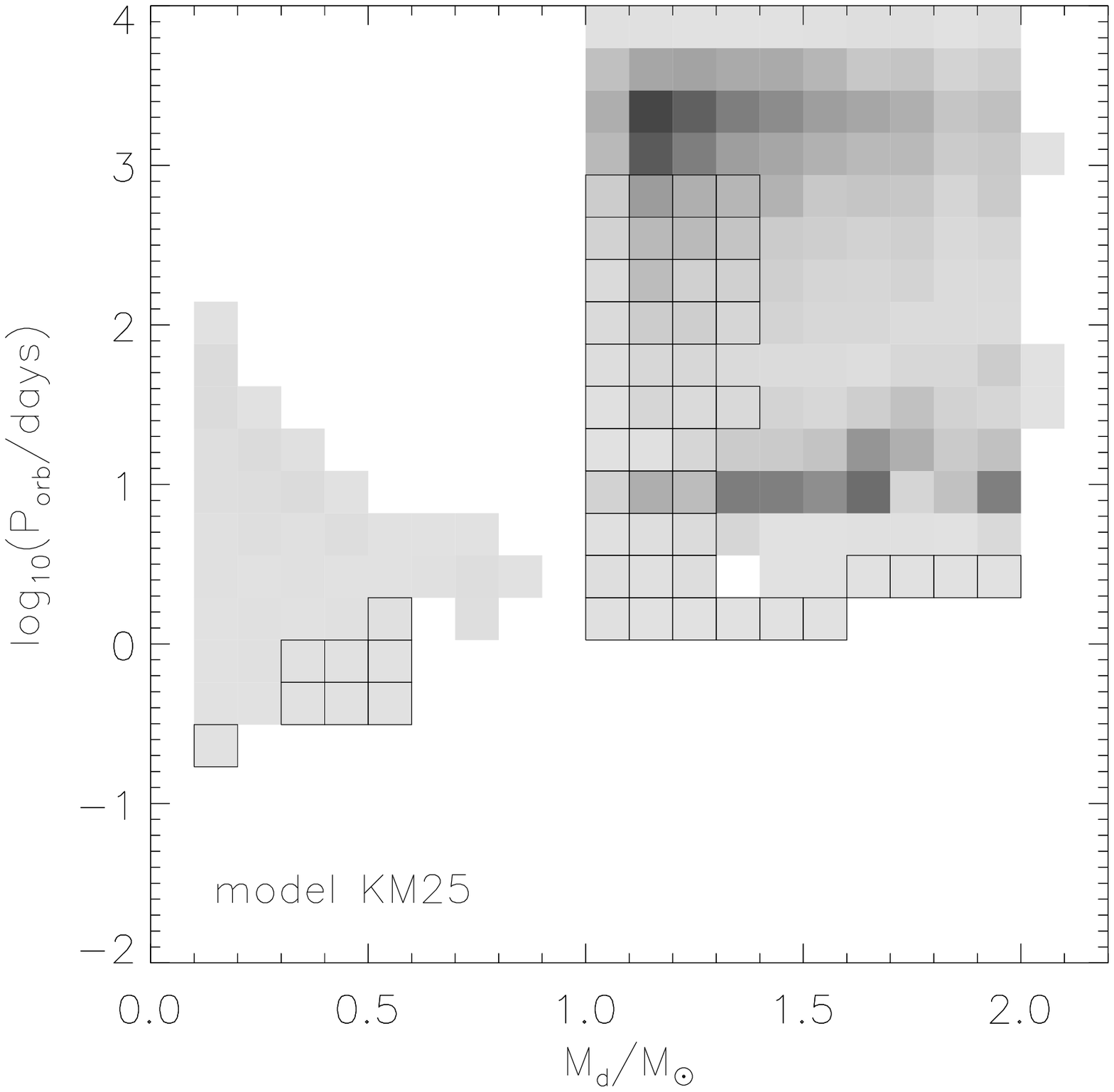}}
\resizebox{5.83cm}{!}{\includegraphics{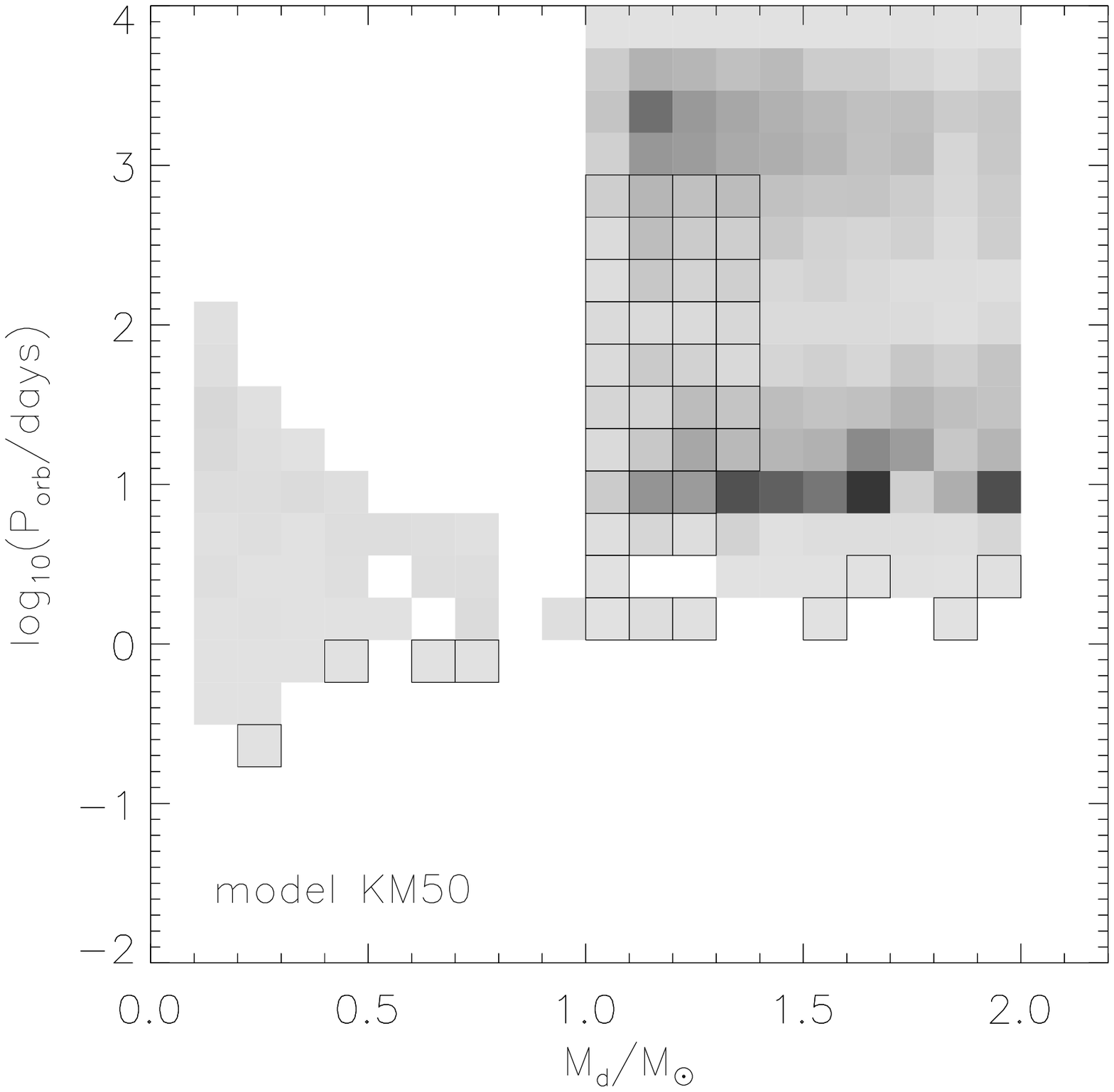}}
\resizebox{5.83cm}{!}{\includegraphics{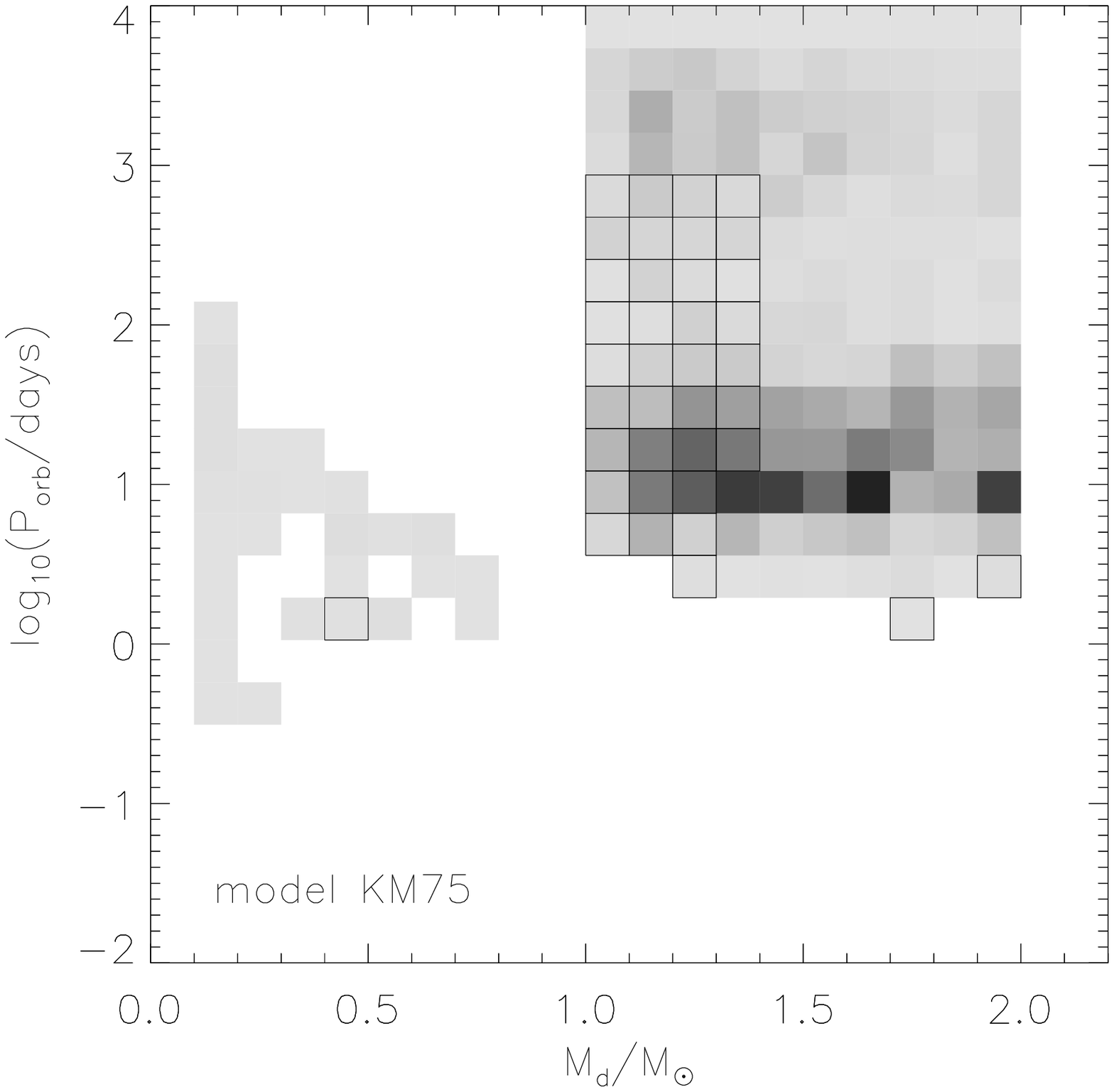}} \\
\resizebox{5.83cm}{!}{\includegraphics{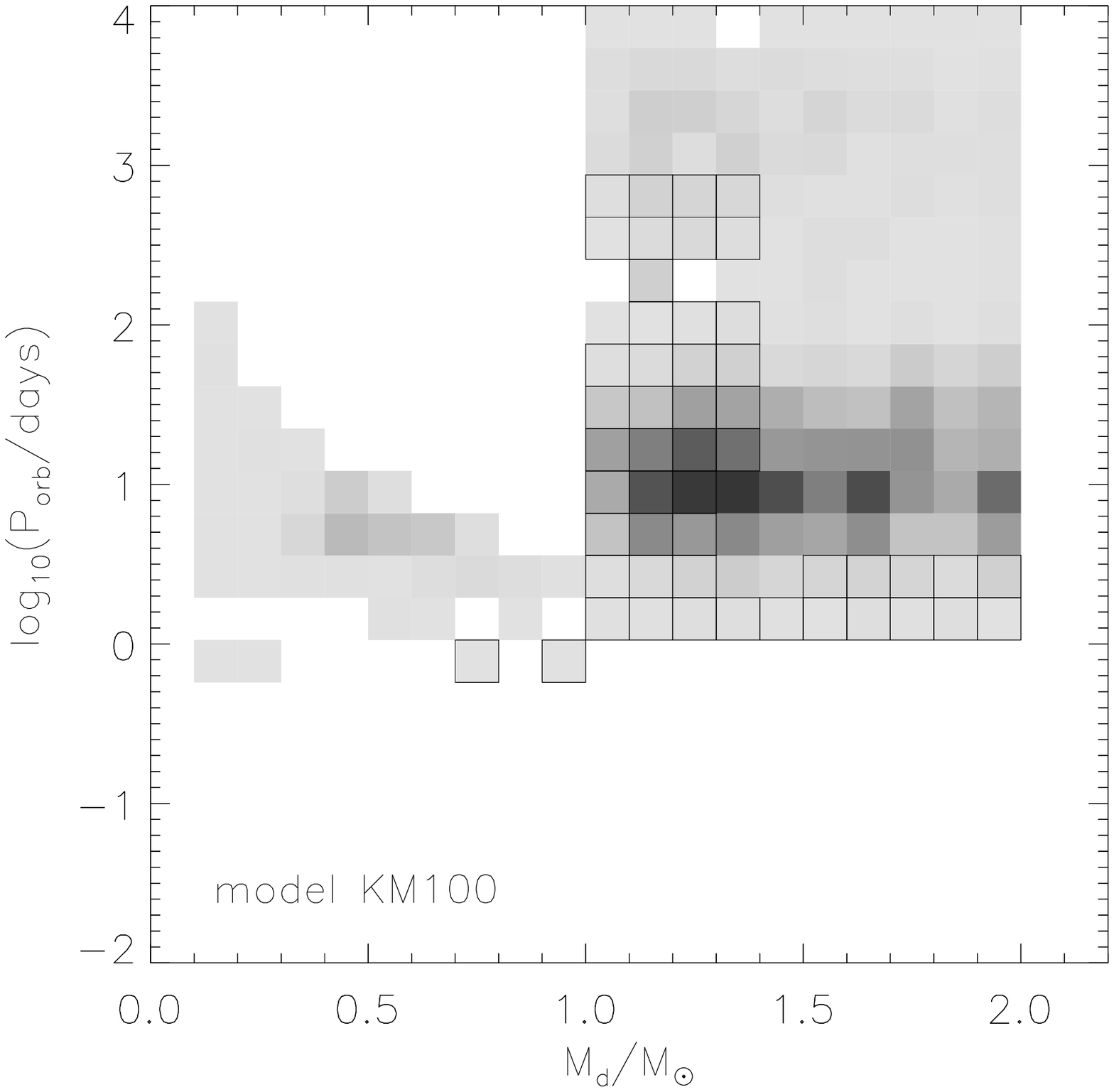}}
\resizebox{5.83cm}{!}{\includegraphics{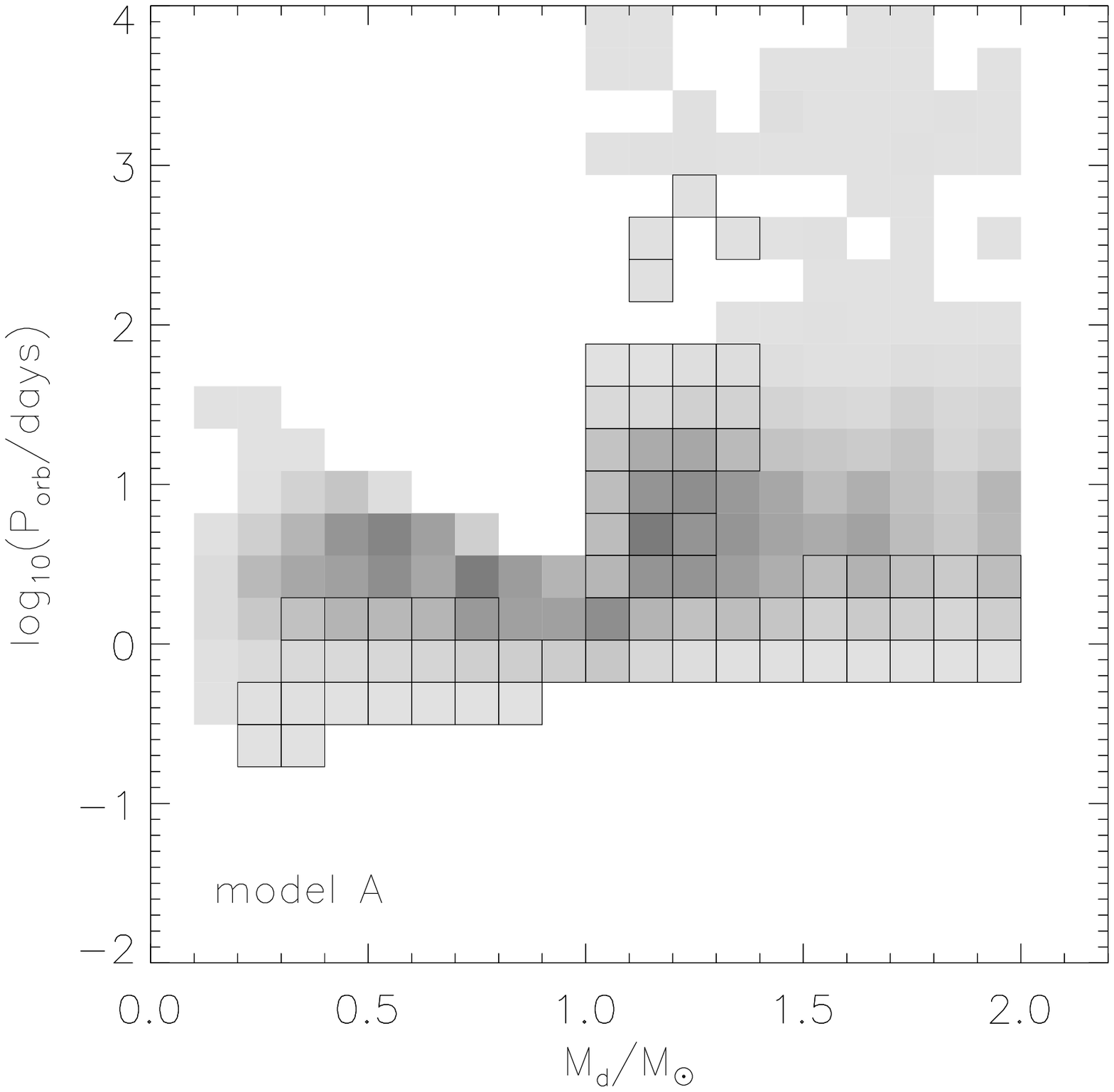}}
\resizebox{5.83cm}{!}{\includegraphics{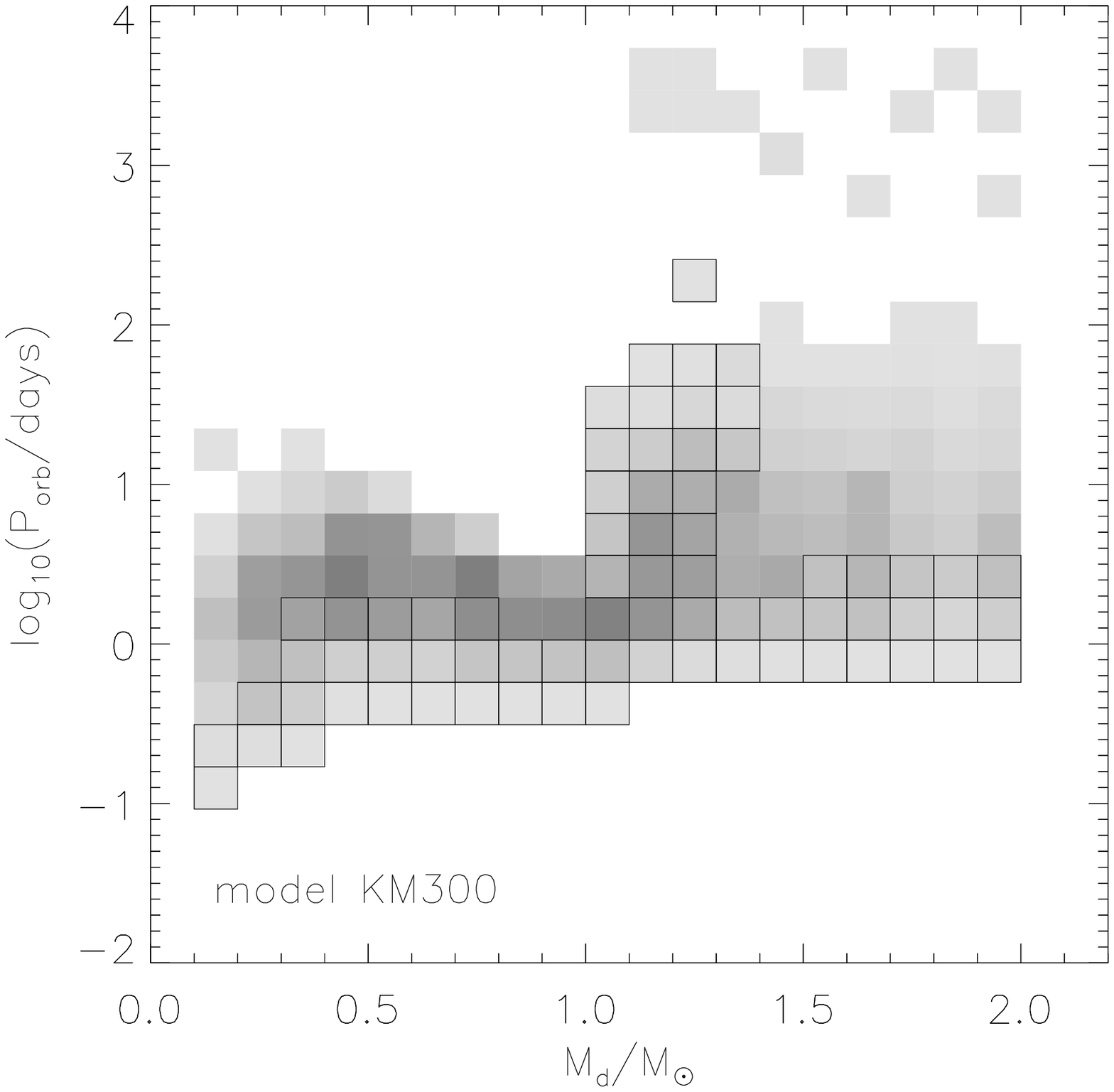}} \\
\resizebox{5.83cm}{!}{\includegraphics{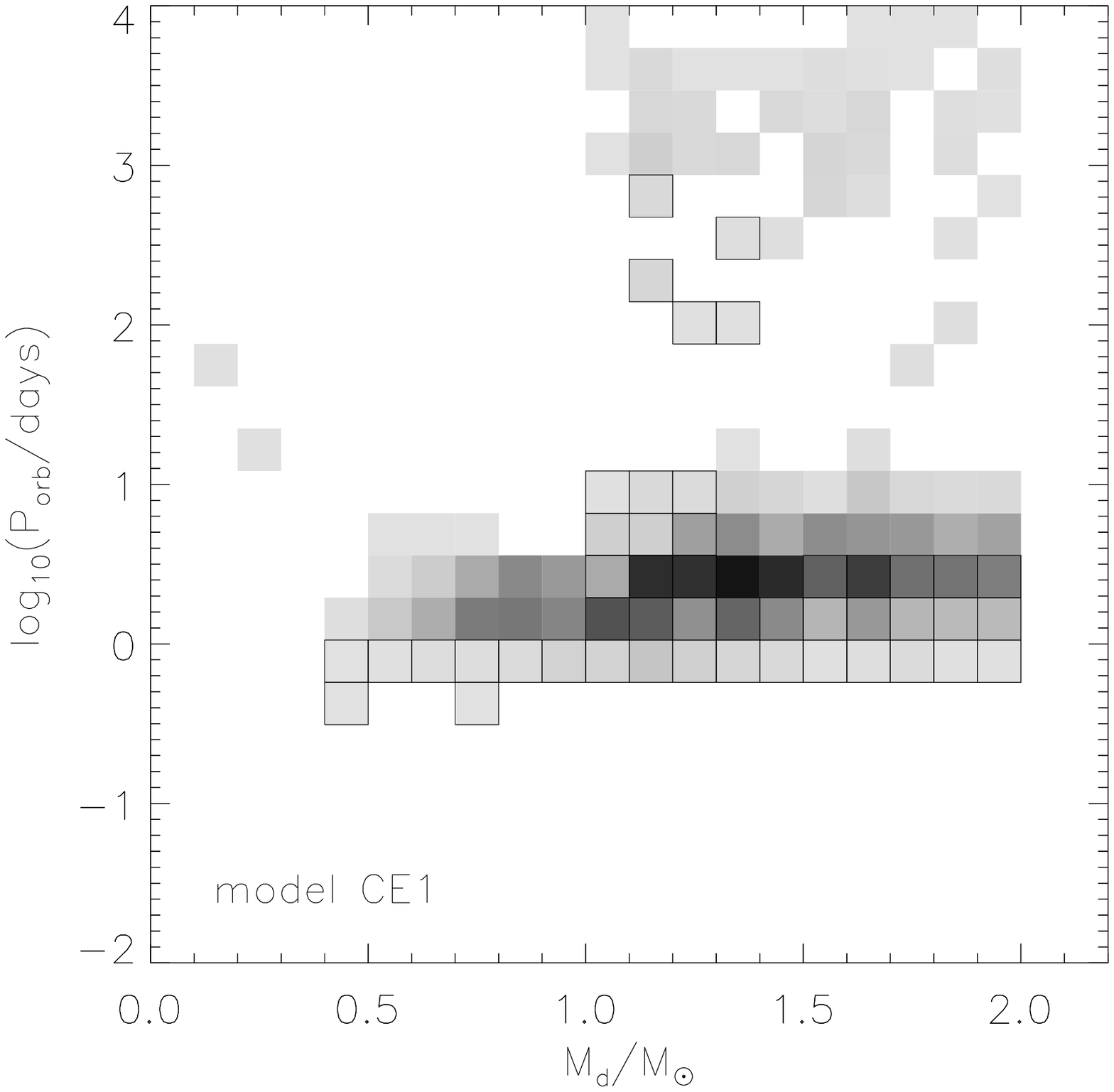}}
\resizebox{5.83cm}{!}{\includegraphics{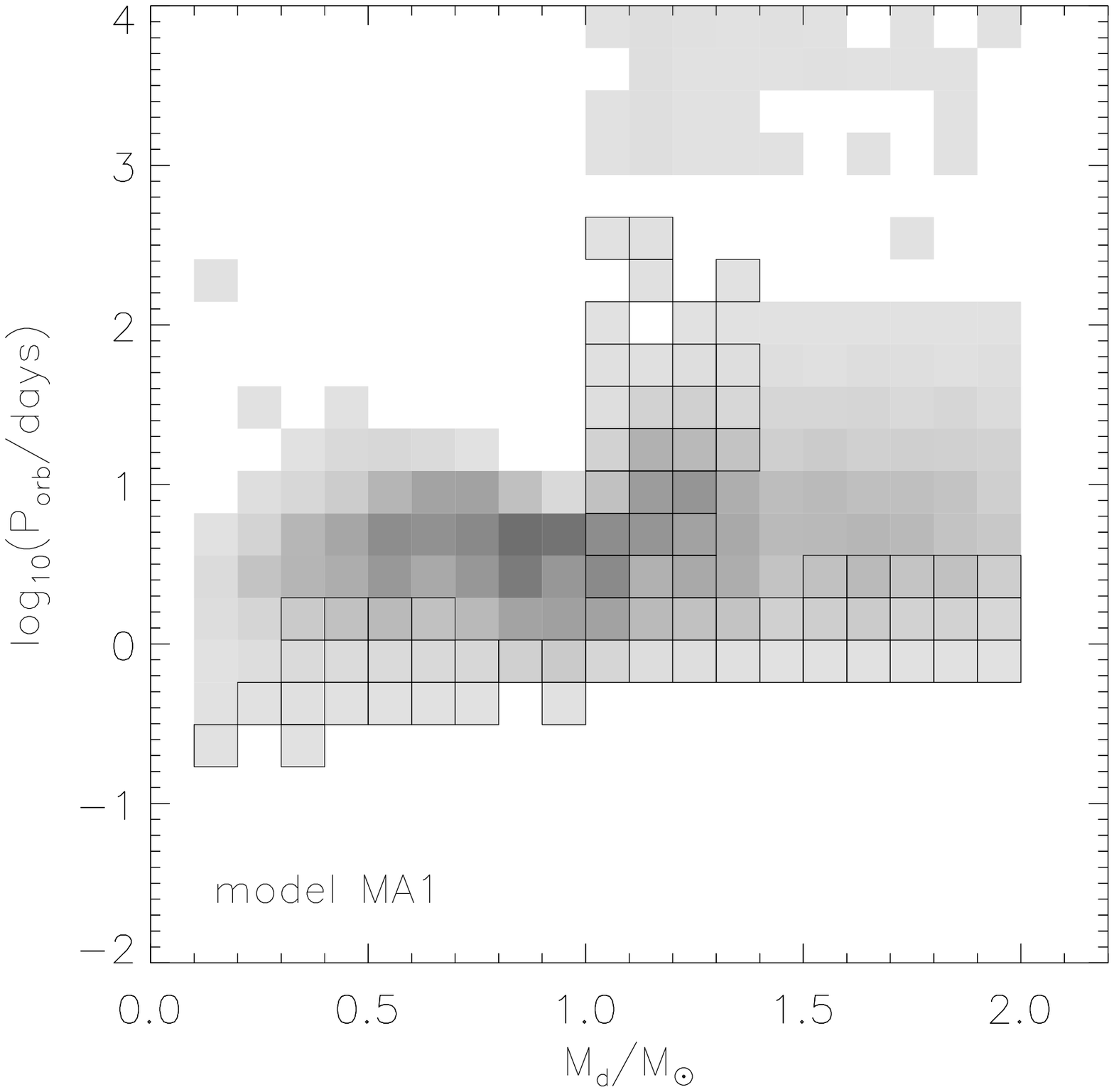}}
\resizebox{5.83cm}{!}{\includegraphics{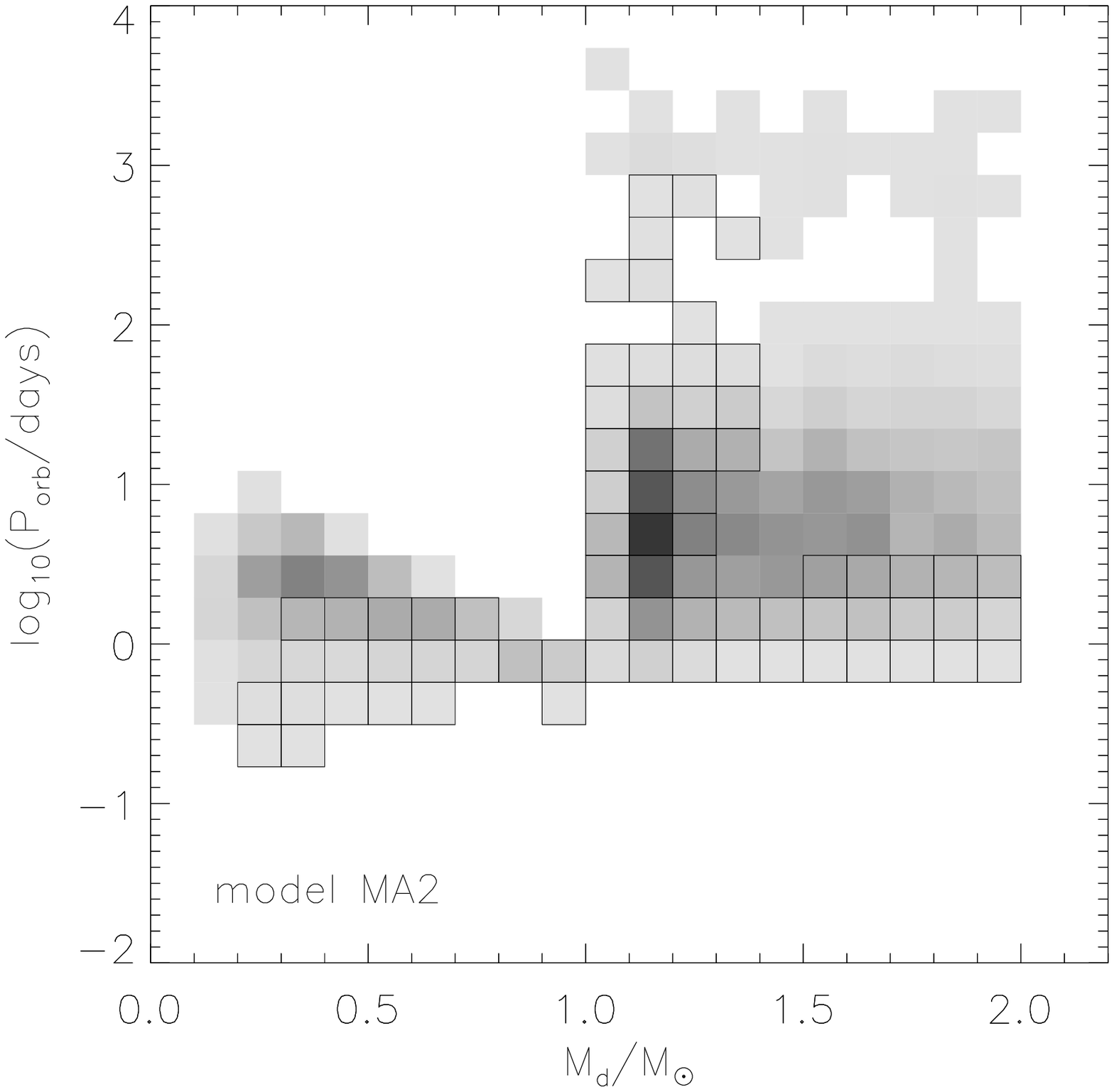}}
\caption{Normalised distribution of pre-LMXBs in the
  $\left(\log M_{\rm d}, \log P_{\rm orb} \right)$-plane. The outlined
  boxes correspond to bins containing pre-LMXBs which actually evolve
  into LMXBs within the imposed age limit of 10\,Gyr.}
\label{MP}
\end{figure*}

The normalised distribution function of pre-LMXBs in the
$\left(\log M_{\rm d}, \log P_{\rm orb} \right)$-plane is displayed
in Fig.~\ref{MP} for
the most illustrative of the population synthesis models listed in
Table~\ref{models}. Bins containing systems that actually evolve
into LMXBs within the imposed age limit of 10\,Gyr are outlined with a
black solid line.

There are three reasons why systems satisfying our pre-LMXB criteria
do not show up in the corresponding LMXB population. Donor stars with
mass $M_{\rm d} \la 1\,M_\odot$ evolve too slowly to fill their Roche
lobe as a result of their nuclear evolution within the imposed age
limit of 10\,Gyr. They can therefore only become LMXBs if their
orbital period is shorter than the bifurcation period of $\sim 1$ day,
which separates diverging from converging systems (Pylyser \& Savonije
1988, 1989). Donor stars with mass $M_{\rm d} \ga 1.4\,M_\odot$ in
systems with $P_{\rm orb} \ga 2.5\,{\rm days}$ on the other hand may
fill their Roche lobe when the donor star ascends the first giant
branch, but the resulting mass transfer is dynamically unstable 
so that a common-envelope phase ensues instead of a LMXB
phase. Systems with $P_{\rm orb} \ga 800\,{\rm days}$, finally, are
too wide for the donor star to fill its Roche lobe prior to the AGB
stage.

The lack of pre-LMXBs in the upper-left corner of each of the plots
presented in Fig.~\ref{MP} is related to our treatment of the coronal
X-ray luminosity of the donor stars. Since stars less massive than
$\sim 1\,M_\odot$ are essentially unevolved and have appreciable
convective envelopes, the adopted upper limit for the coronal X-ray
luminosity is proportional to the square of the zero-age main sequence
radii [Eq.~(\ref{Lcor})]. The decrease of the latter with decreasing
mass then implies that binaries with lower-mass donor stars can
satisfy the criterion $L_{\rm acc} > 10\,L_{\rm cor}$ up to longer
orbital periods. The exact position of the lower border of the wedge
in the upper-left corner of the figures introduced by this criterion
depends on the wind-velocity parameter $\beta_{\rm w}$ and on the
assumptions made for the determination of the coronal X-ray luminosity
$L_{\rm cor}$. Increasing the wind-velocity parameter (model MA2)
yields lower mass-accretion rates and thus lower accretion
luminosities. Correspondingly, the bottom of the wedge shifts to lower
orbital periods. Adopting a constant upper limit for the coronal X-ray
luminosity (models COR1 and COR2, not shown) instead of
Eq.~(\ref{Lcor}) renders the border of the wedge introduced by the
$L_{\rm acc} > 10\,L_{\rm cor}$ criterion independent of the mass of
the donor star.

Stars more massive than $\sim 1\,M_\odot$, on the other hand, may
evolve away from the ZAMS within the imposed age limit of 10\,Gyr.
However, the deep convective envelopes and large radii 
developed on the giant
branch yield high upper limits for the coronal X-ray luminosity so
that only few wind-accreting neutron stars manage to top this limit
with their accretion luminosity. Most of the pre-LMXBs with $M_{\rm d}
\ga 1\,M_\odot$ therefore also have main-sequence donor stars. The
rather sharp vertical border of the wedge in the upper-left corner of
the figures at $M_{\rm d} \approx 1\,M_\odot$ is then related to the
assumed lower limit $M_{\rm conv}=0.01\,M_{\rm d}$ for the mass of the
convective envelope of coronally active stars. Main-sequence stars
more massive than $\sim 1\,M_\odot$ have smaller or no convective
envelopes and are thus assumed to have negligible coronal X-ray
luminosities. The pre-LMXB luminosity criterion therefore reduces
to $L_{\rm acc} > 10^{-6}\,L_\odot$, which is easily satisfied. If
the minimum convective envelope mass necessary for X-ray activity is
increased to, e.g., $M_{\rm conv} = 0.1\, M_{\rm d}$ (model COR3,
not shown), the sharp vertical border of the wedge shifts to
$M_{\rm d} \simeq 0.8\,M_\odot$.

The group of systems with orbital periods longer than $\sim 1000$ days
and donor star masses between 1.0 and 1.5\,$M_\odot$ in the low
kick-velocity dispersion models shown in Fig.~\ref{MP} corresponds to
systems with initial orbital separations wide enough to avoid any kind
of Roche-lobe overflow prior to the formation of the neutron
star. These systems easily survive the supernova explosion of the
neutron star's progenitor when the average kick velocity is small
compared to the relative orbital velocity of the component stars, but
generally get disrupted for higher average kick velocities.

The narrow horizontal ridge of systems near $P_{\rm orb} \approx 10$
days in the low kick-velocity dispersion models ($\sigma_{\rm kick}
\la 100$ km/s) is associated with systems that spiralled-in during a
common-envelope phase triggered by Roche-lobe overflow from the
neutron star's progenitor prior to its supernova explosion. The range
of orbital periods available for this evolutionary channel is limited
by the requirements that the pre-common-envelope orbit must be small
enough to initiate Roche-lobe overflow prior to the supernova
explosion and wide enough to provide enough orbital energy to expel
the common envelope before the core of the Roche-lobe overflowing star
merges with the spiralling-in companion. After the expulsion of the
envelope, the change in the orbital configuration caused by the
supernova explosion depends on the amount of mass lost from the system
and on the kick velocity imparted to the neutron star at birth.  For a
detailed account of the effects of asymmetric supernova explosions on
the orbital characteristics of binary stars, we refer to Kalogera
(1996).

In the case of small supernova kick-velocity dispersions, the effect
of the mass loss dominates over the effect of the kick. The small
spread of orbital periods available for the common-envelope scenario
is then preserved by the one-to-one relation between the
post-supernova orbital period and the mass lost from the system
(e.g. Verbunt 1993). Furthermore, since binaries in which more than
half of the total mass is lost from the system become unbound after a
symmetric supernova explosion, the probability of survival is larger
for binaries with more massive donor stars.

In the case of higher supernova kick-velocity dispersions the effect
of the kick dominates over the effect of the mass loss. This
facilitates the survival of binaries with lower-mass donor stars and
increases the spread of the post-supernova orbital periods. The
horizontal ridge around $P_{\rm orb} \approx 10$ days therefore widens
and shifts to lower donor star masses.

The common-envelope ejection efficiency parameter $\alpha_{\rm CE}$
only affects systems with orbital periods less than $\sim 1000$
days. A decrease of $\alpha_{\rm CE}$ (model CE1) yields smaller
orbital separations at the end of the common-envelope phase and thus
shifts the horizontal ridge near $P_{\rm orb} \approx 10$ days to
lower orbital periods. This effectively widens the gap between the
long- and short-period systems. Increasing the common-envelope
efficiency $\alpha_{\rm CE}$ has the opposite effect.

We conclude this section by noting that the position of the
high-density areas in Fig.~\ref{MP} also depends on the adopted
initial mass-ratio distribution. Mass ratio distributions favouring
lower initial mass ratios (e.g. $n(q)=1/q$) shift the high-density
areas to lower donor star masses, while mass ratio distributions
favouring higher initial mass ratios (e.g. $n(q)=q$) shift the
high-density areas to higher donor star masses.

\section{X-ray luminosities}

\begin{figure*}
\resizebox{5.83cm}{!}{\includegraphics{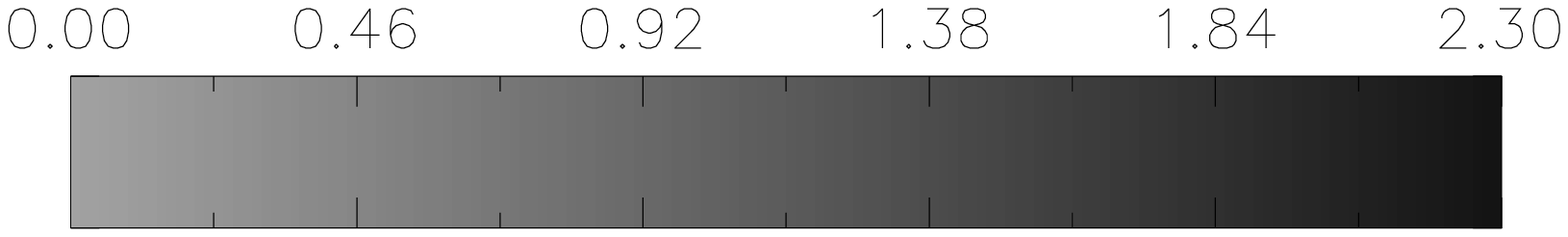}} \\
\resizebox{5.83cm}{!}{\includegraphics{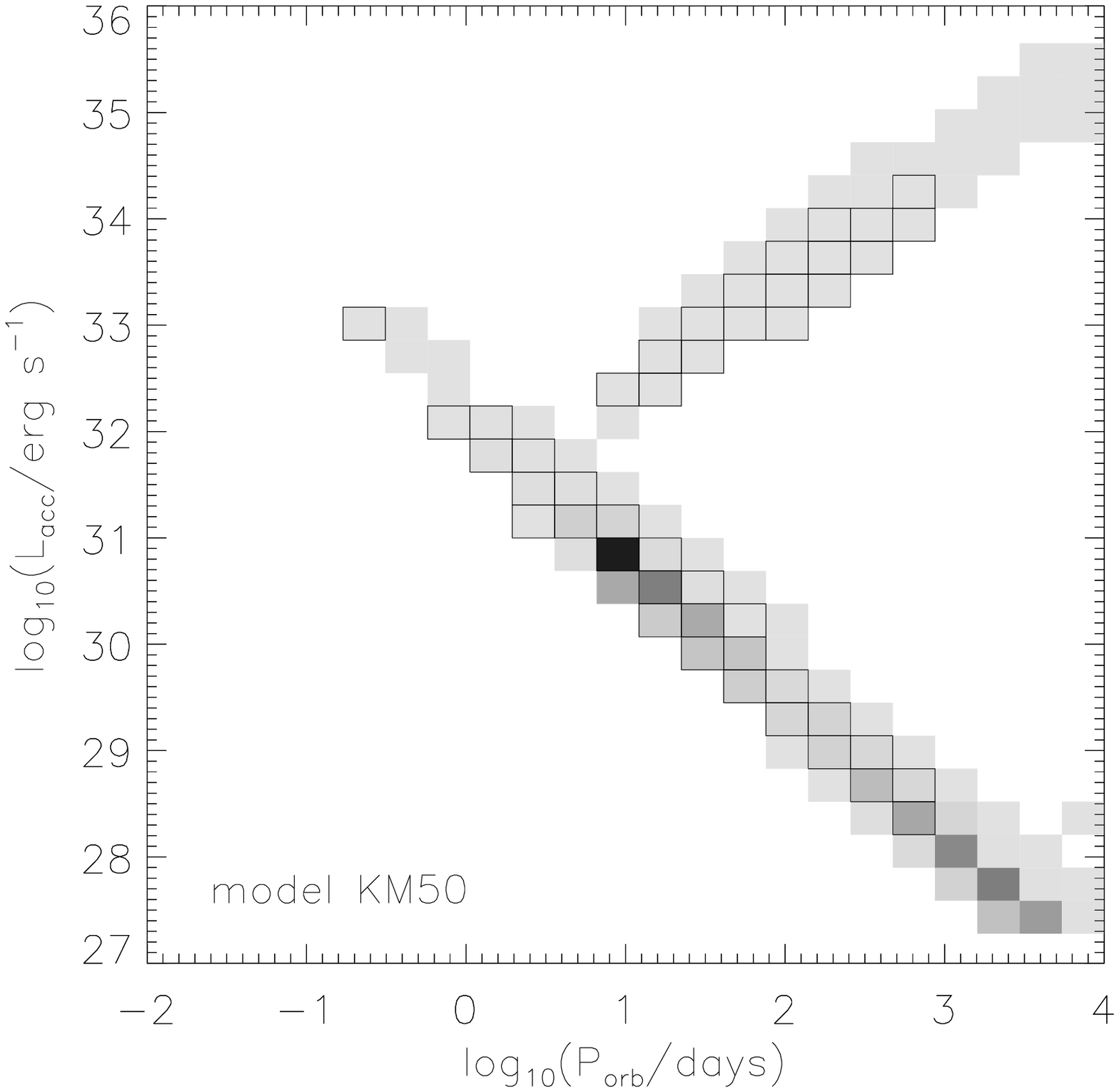}}
\resizebox{5.83cm}{!}{\includegraphics{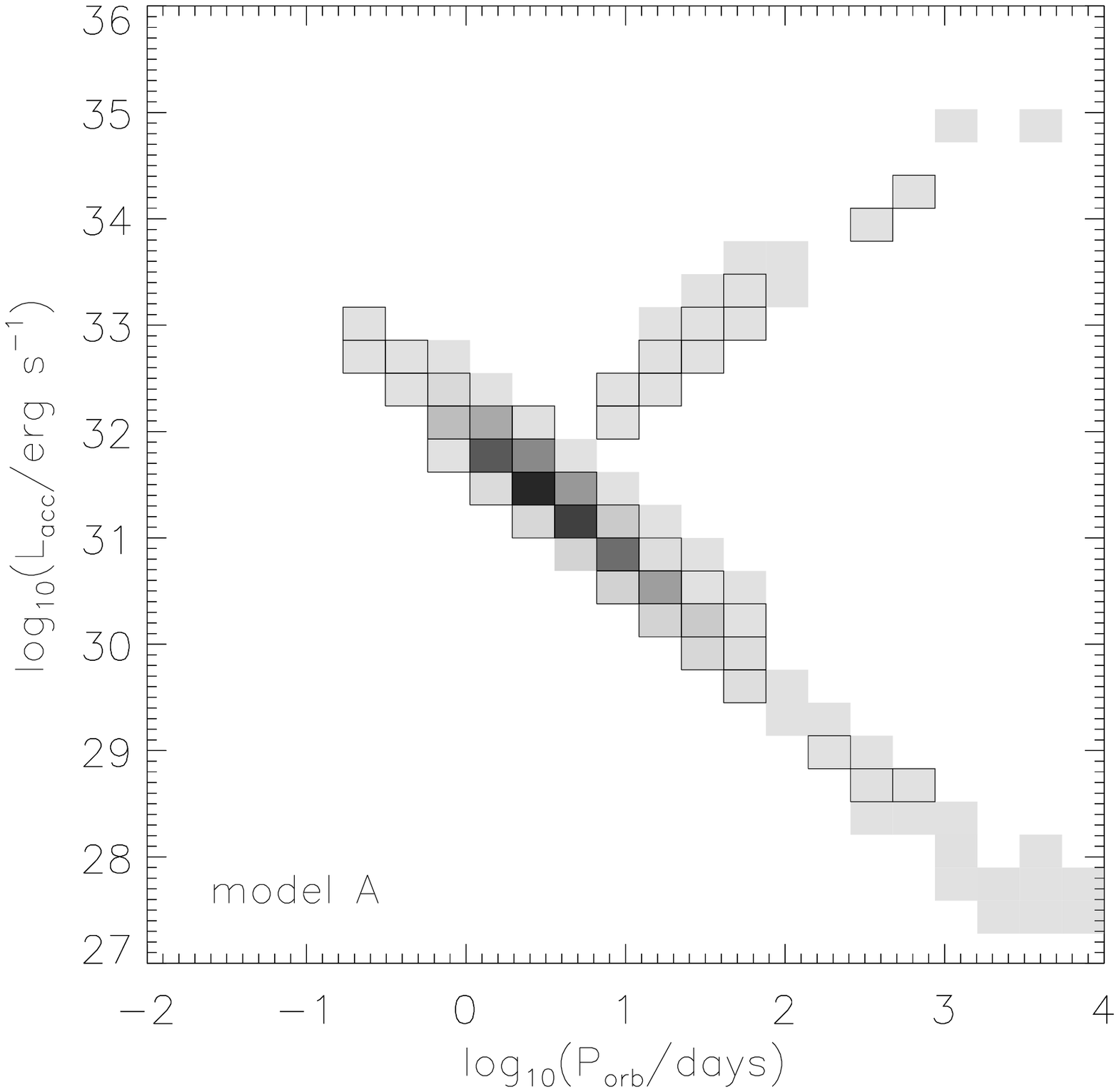}}
\resizebox{5.83cm}{!}{\includegraphics{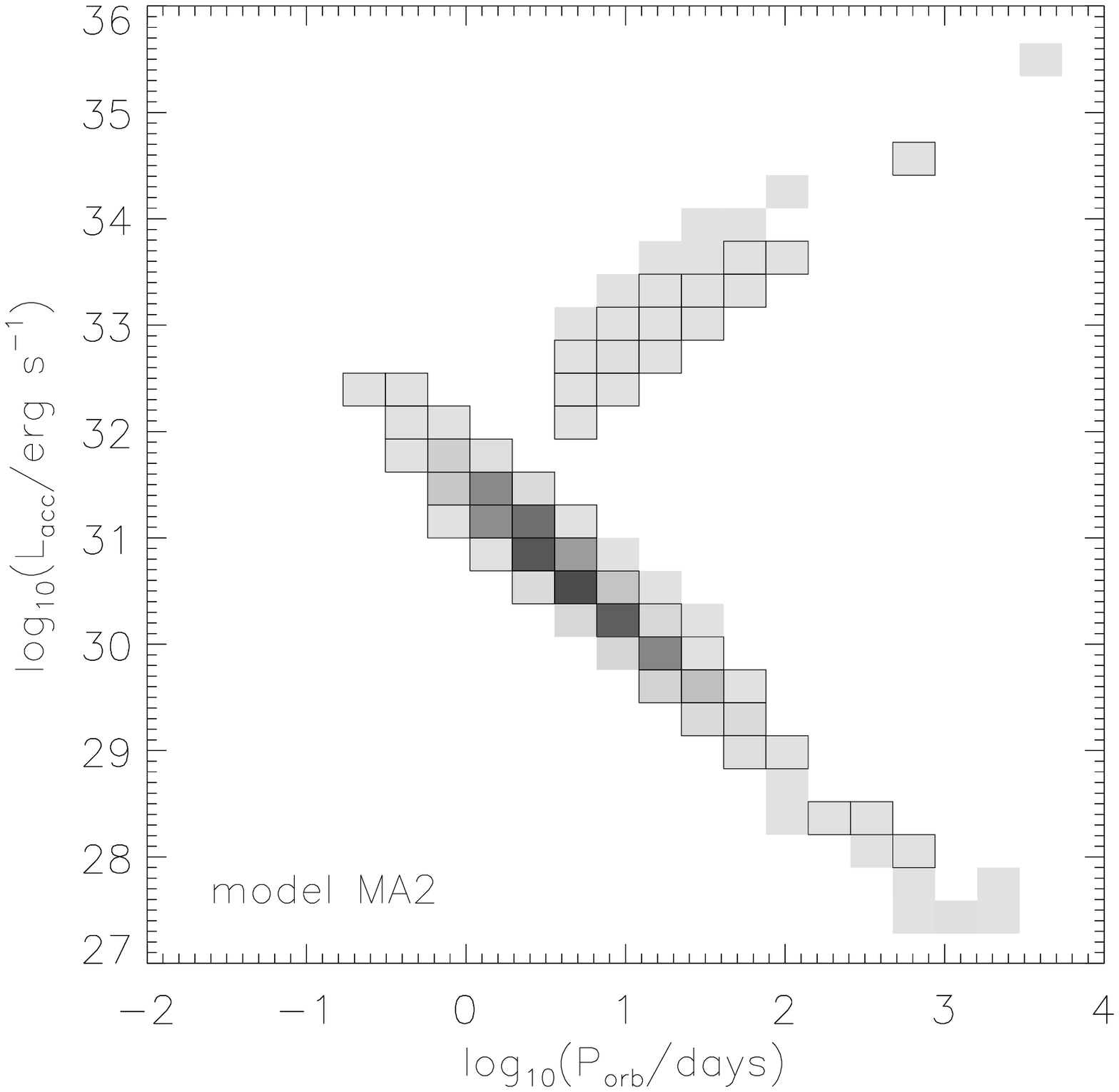}}
\caption{Normalised distribution of pre-LMXBs in the
  $\left(\log P_{\rm orb}, \log L_{\rm acc} \right)$-plane. The
  outlined boxes correspond to bins containing pre-LMXBs which
  actually evolve into LMXBs within the imposed age limit of 10\,Gyr.}
\label{PL}
\end{figure*}

\begin{figure*}
\resizebox{5.83cm}{!}{\includegraphics{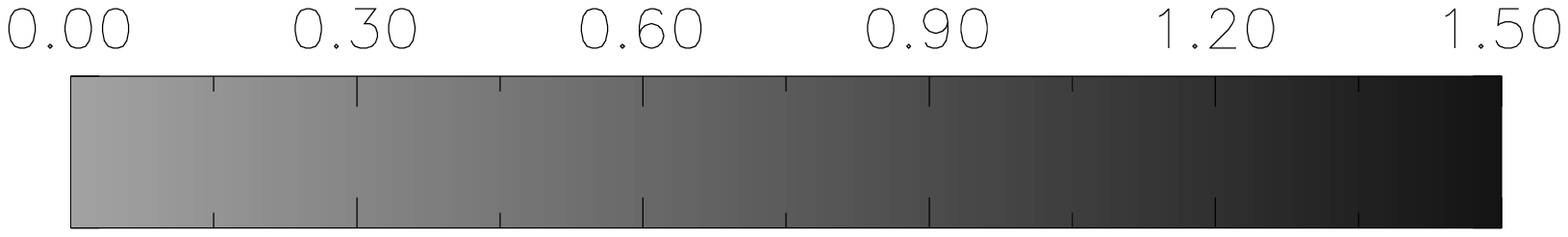}} \\
\resizebox{5.83cm}{!}{\includegraphics{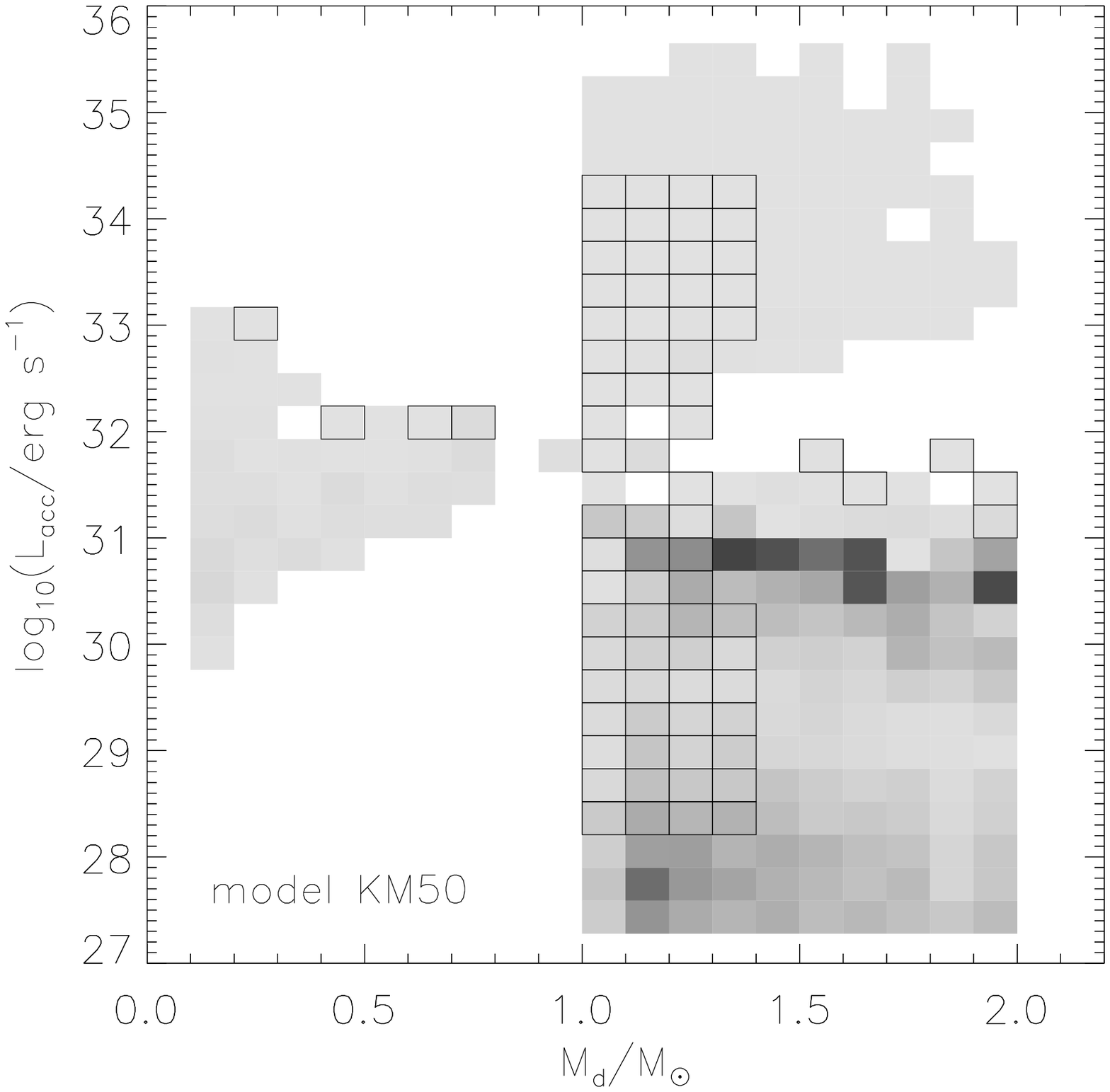}}
\resizebox{5.83cm}{!}{\includegraphics{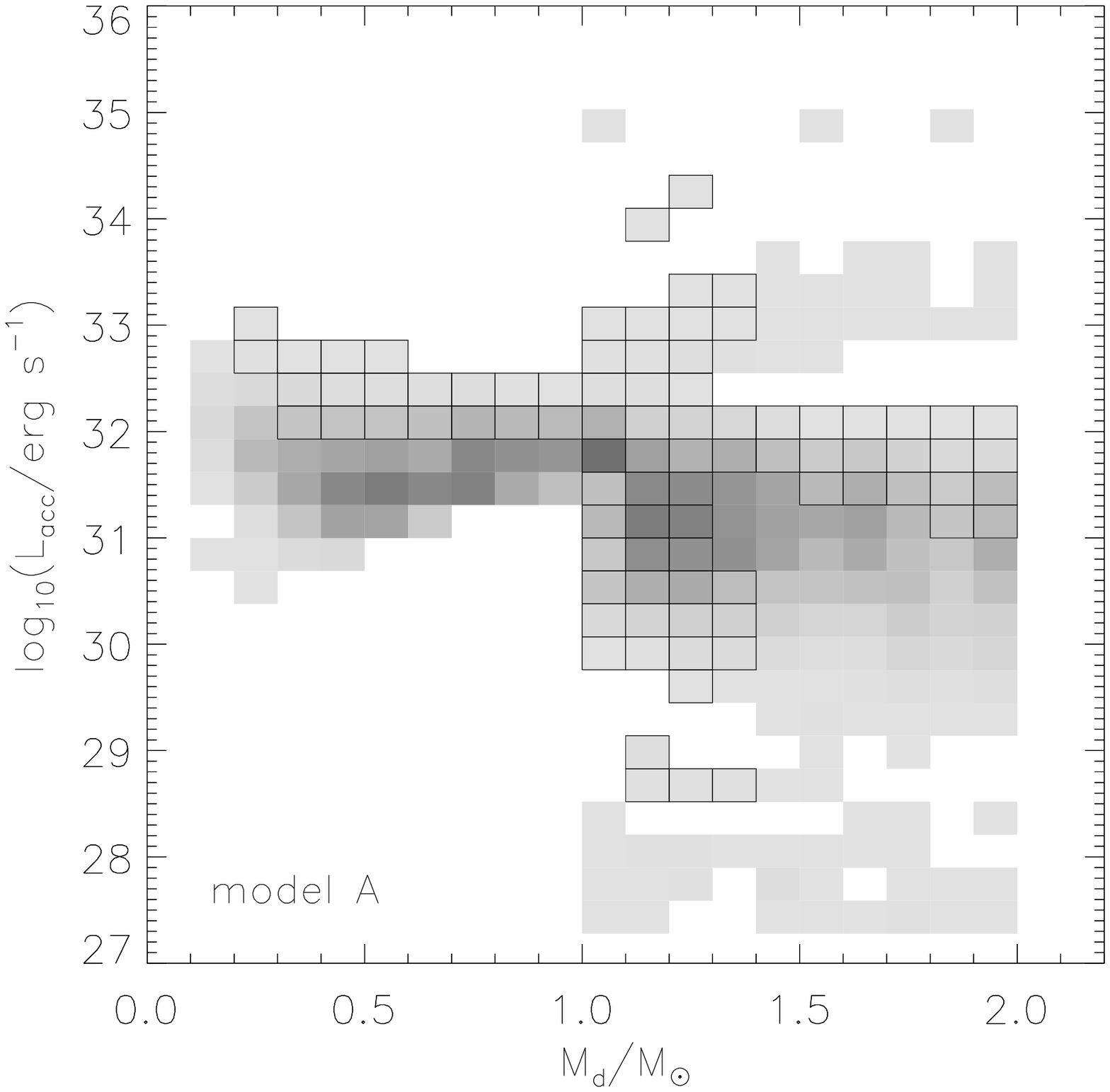}}
\resizebox{5.83cm}{!}{\includegraphics{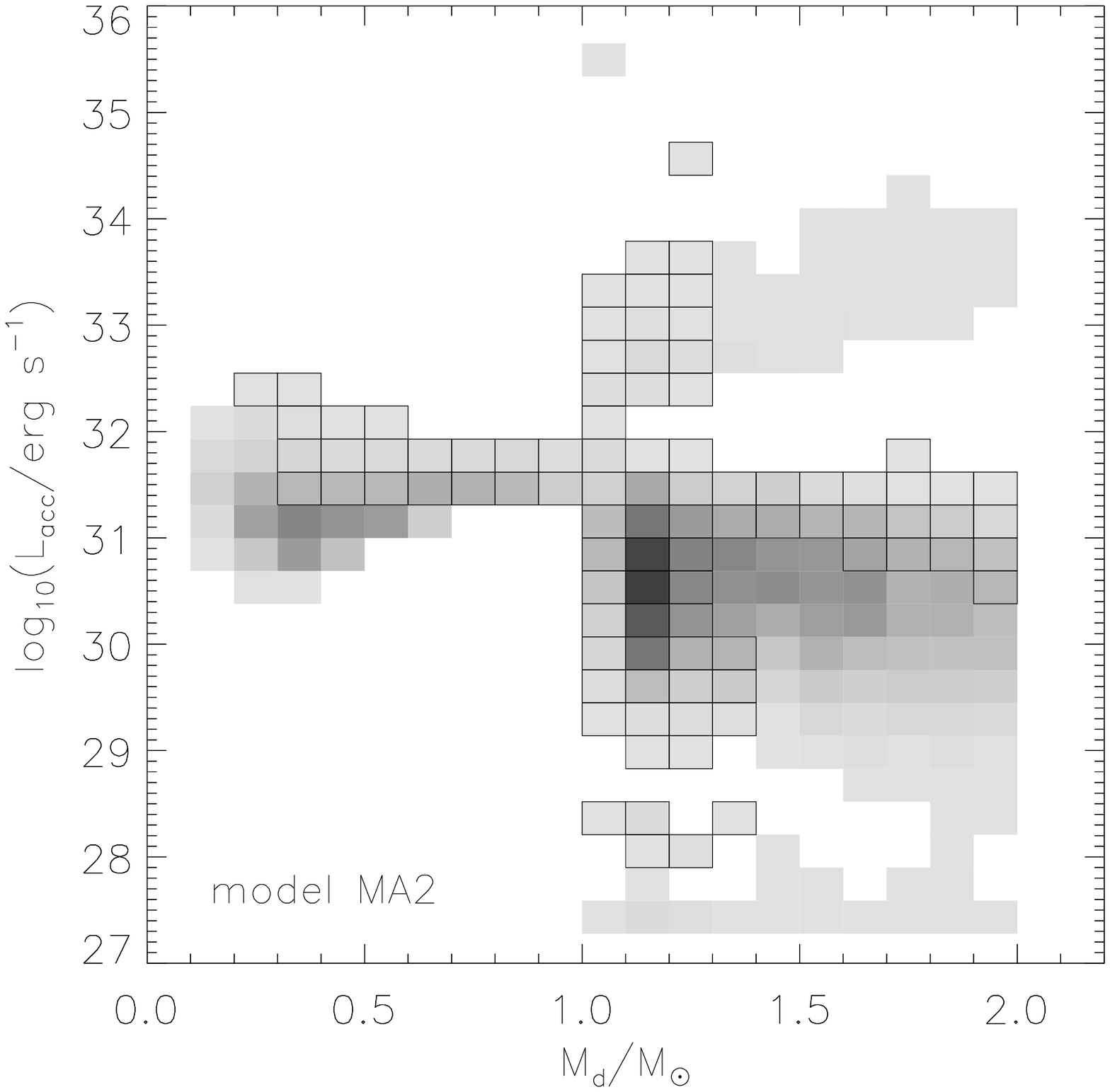}}
\caption{Normalised distribution of pre-LMXBs in the
  $\left(\log M_{\rm d},\log L_{\rm acc} \right)$-plane. The outlined
  boxes correspond to bins containing pre-LMXBs which actually evolve
  into LMXBs within the imposed age limit of 10\,Gyr.}
\label{ML}
\end{figure*}

The normalised distribution functions of pre-LMXBs in the
$\left(\log P_{\rm orb}, \log L_{\rm acc} \right)$- and the
$\left(\log M_{\rm d},\log L_{\rm acc} \right)$-plane are displayed in
Figs.~\ref{PL} and~\ref{ML}, respectively. Since the qualitative
differences between the models listed in Table~\ref{models} are
small, we limit the figures to models~KM50, A, and MA2.

The bulk of the systems shown in Fig.~\ref{PL} have logarithmic
accretion luminosities that are linearly correlated with the logarithm
of the orbital period. The slope of -4/3 arises from Kepler's third
law and the factor $1/a^2$ in Eq. (\ref{Macc2}) for the mean
mass-accretion rate $\dot{M}_{\rm acc}$. Since most of the systems
following the correlation have main-sequence donor stars, the small
spread around the correlation is the result of the limited range of
donor star masses and radii and the adopted constant wind mass-loss
rate of $10^{-13}\,M_\odot$ per year [see Eq.~(\ref{Macc2})]. In the
case of low supernova kick-velocity dispersions, the two distinct
higher-density regions along the $L_{\rm acc} \propto P_{\rm
orb}^{-4/3}$ relation correspond to the two groups of systems
identified in the low kick-velocity dispersion models shown in
Fig.~\ref{MP}. The high accretion luminosities around $10^{31}$ erg/s
stem from systems on the horizontal ridge around $P_{\rm orb} \approx
10$ days, while the lower accretion luminosities around $10^{28}$
erg/s stem from the long-period systems with $P_{\rm orb} \ga 1000$
days. For higher kick-velocity dispersions, only the first group of
systems survives. The accretion luminosities of the pre-LMXBs on the
$L_{\rm acc} \propto P_{\rm orb}^{-4/3}$ furthermore decrease with
increasing values of the wind-velocity parameter $\beta_{\rm w}$.

The branch of systems located to the right of the $L_{\rm acc} \propto
P_{\rm orb}^{-4/3}$ correlation corresponds to pre-LMXBs with giant
branch donor stars. Most of these already appeared as pre-LMXBs when
the donor star was still on the main-sequence, but the growth of the
convective envelope and the associated increase in the coronal X-ray
activity during the post-main-sequence evolution temporarily concealed
the X-ray accretion luminosity emitted by the wind-accreting neutron
star. The functional form of the mean mass-accretion rate
$\dot{M}_{\rm acc}$ [Eq. (\ref{Macc2})] implies that systems with
longer orbital periods require more evolved donor stars, i.e. stars
with larger radii and higher wind mass-loss rates, to satisfy the
imposed $L_{\rm acc}>10\,L_{\rm cor}$ criterion. However, the larger
radii in turn imply a larger upper limit for the coronal X-ray
luminosity $L_{\rm cor}$ [Eq. (\ref{Lcor})], so that at the time of
reappearance the $10\,L_{\rm cor}$ threshold itself will be higher for
systems with wider orbital separations.  Hence, the accretion
luminosity at which the systems reappear as pre-LMXBs increases with
increasing orbital periods.  Their reappearance also depends
critically on whether or not the accretion luminosity increases faster
than the coronal X-ray luminosity.

The position of the giant-branch stars in the $\left(\log P_{\rm
orb}, \log L_{\rm acc} \right)$-plane clearly depends on the
assumptions adopted for the determination of $L_{\rm acc}$ and $L_{\rm
cor}$. Since higher wind velocities yield lower mass-accretion rates,
an increase in the wind-velocity parameter $\beta_{\rm w}$ as in
model~MA2 requires even more evolved donor stars to satisfy the
$L_{\rm acc}>10\,L_{\rm cor}$ criterion than in model~A. The
reappearance threshold of these systems consequently moves to higher
accretion luminosities. In the case of a constant coronal X-ray
luminosity (models COR1 and COR2, not shown), the accretion luminosity
at which systems with giant branch donor stars reappear as pre-LMXBs
is independent of the orbital period. The temporary concealment of the
accretion luminosity probably does not occur for the longer period
systems ($P_{\rm orb} \ga 100$ days) in which tidal forces are too
weak to keep the donor star's rotation synchronised with the orbital
motion of the companion. The decrease of the donor star's rotational
angular velocity during its post-main-sequence evolution is here more
likely to be accompanied by a decrease in the coronal X-ray activity
so that the wind-accretion luminosity in these binaries has no problem
topping the coronal X-ray luminosity at any time after they first
appear as a pre-LMXB.

The linear correlation between $\log L_{\rm acc}$ and $\log P_{\rm
orb}$ for the bulk of the systems presented in Fig.~\ref{PL}
essentially makes Fig.~\ref{ML} a mirror image of Fig.~\ref{MP}. The
low kick-velocity dispersion models hence again show two distinct
groups of systems at high and low accretion luminosities, which are
associated with the horizontal ridge and the long-period systems
observed in Fig.~\ref{MP}. In our standard model (model~A), even
systems with donor stars less massive than $1.0\,M_\odot$ are seen to
give rise to accretion luminosities up to $10^{32}$ erg/s. An increase
in the wind-velocity parameter $\beta_{\rm w}$ shifts all systems to
lower accretion luminosities.

\begin{figure*}
\resizebox{5.83cm}{!}{\includegraphics{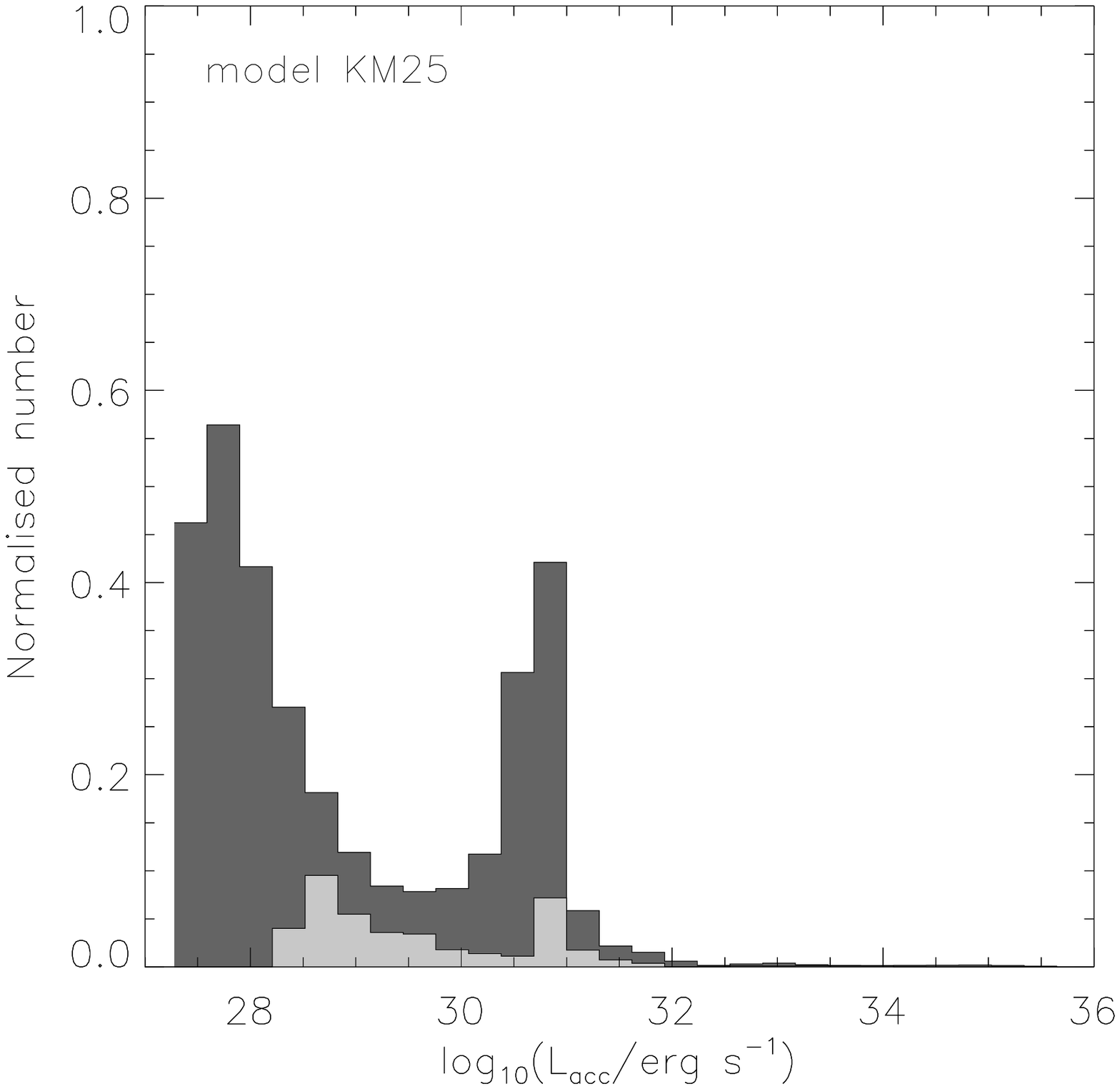}}
\resizebox{5.83cm}{!}{\includegraphics{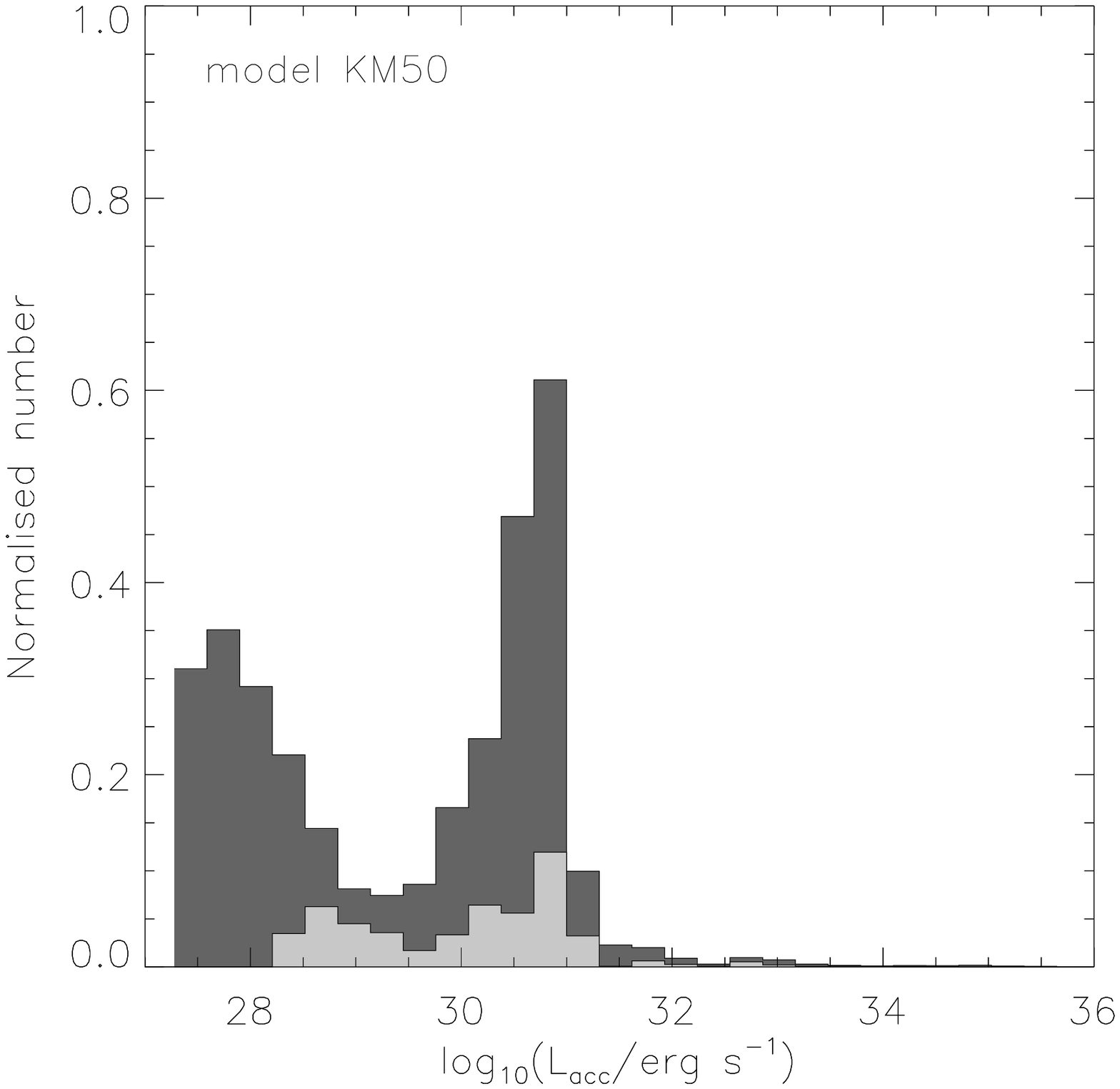}}
\resizebox{5.83cm}{!}{\includegraphics{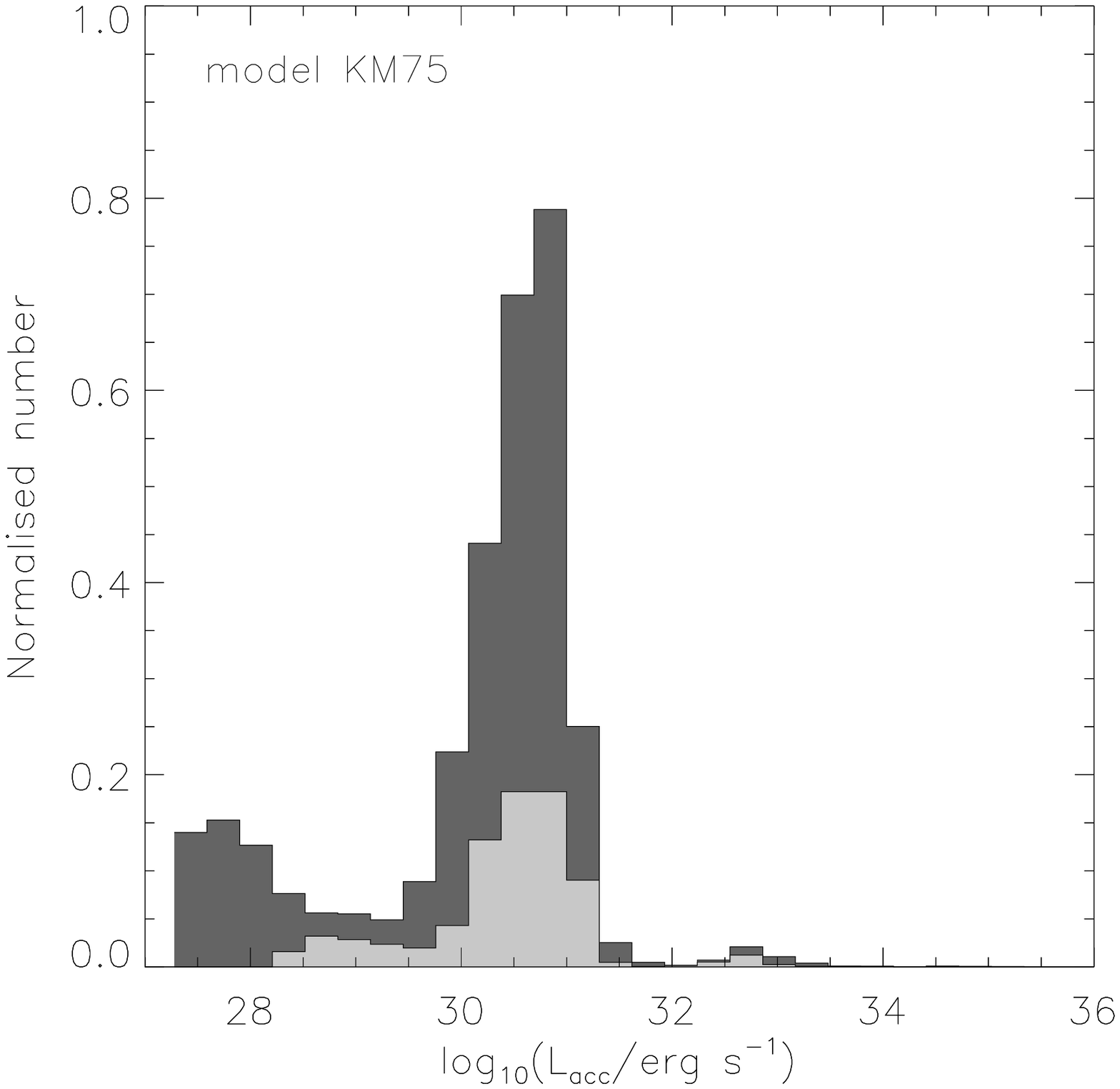}} \\
\resizebox{5.83cm}{!}{\includegraphics{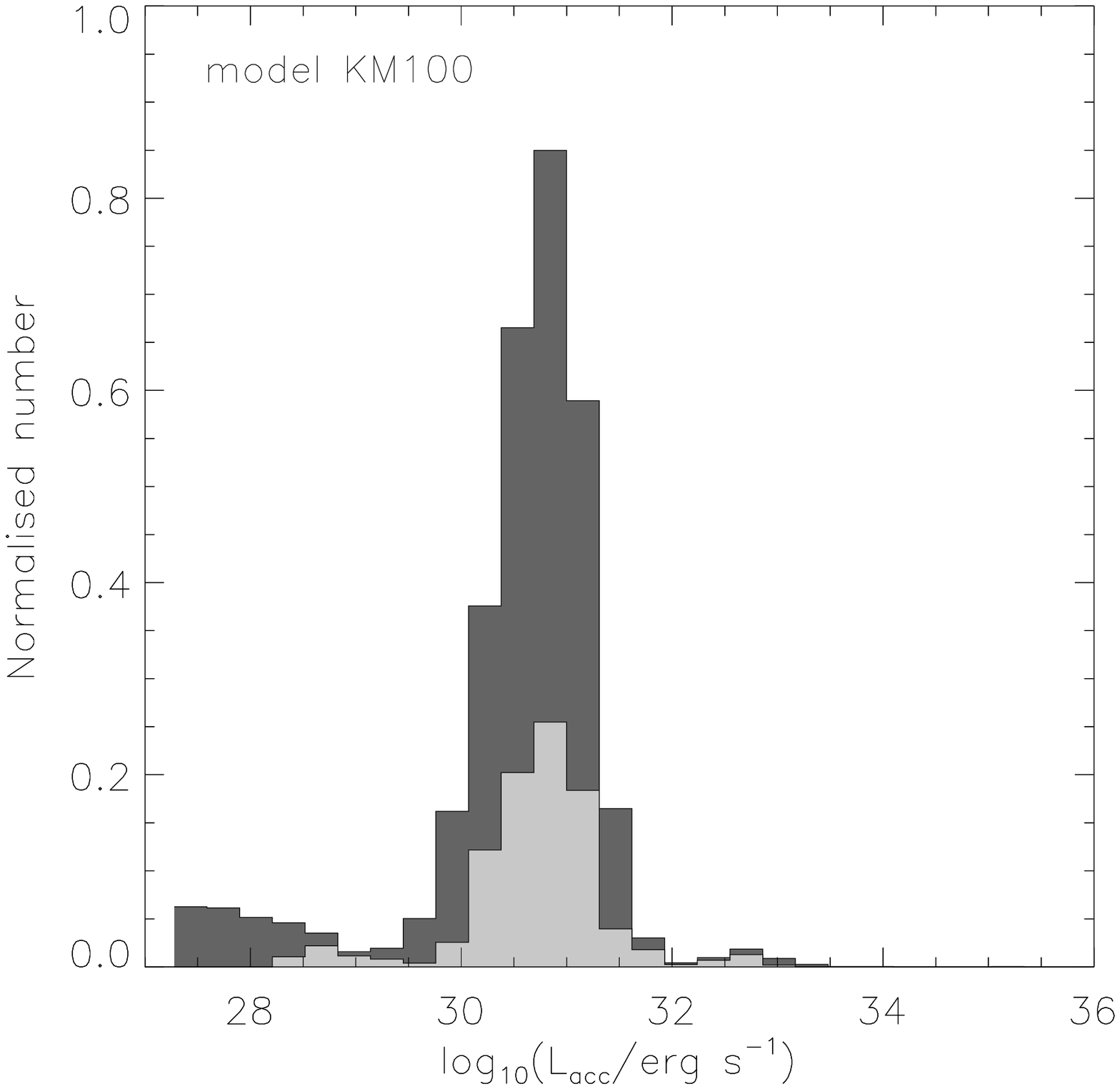}}
\resizebox{5.83cm}{!}{\includegraphics{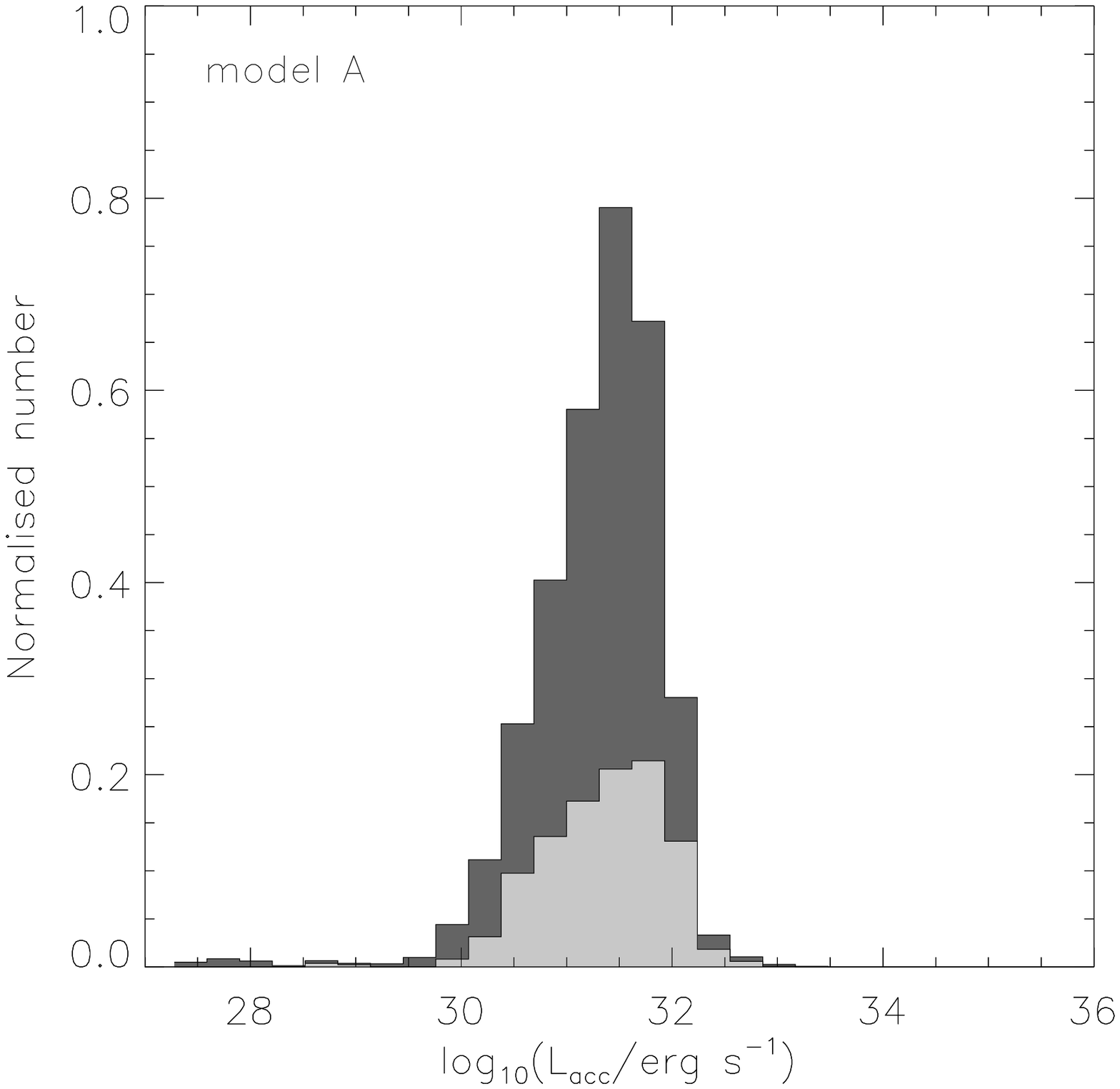}}
\resizebox{5.83cm}{!}{\includegraphics{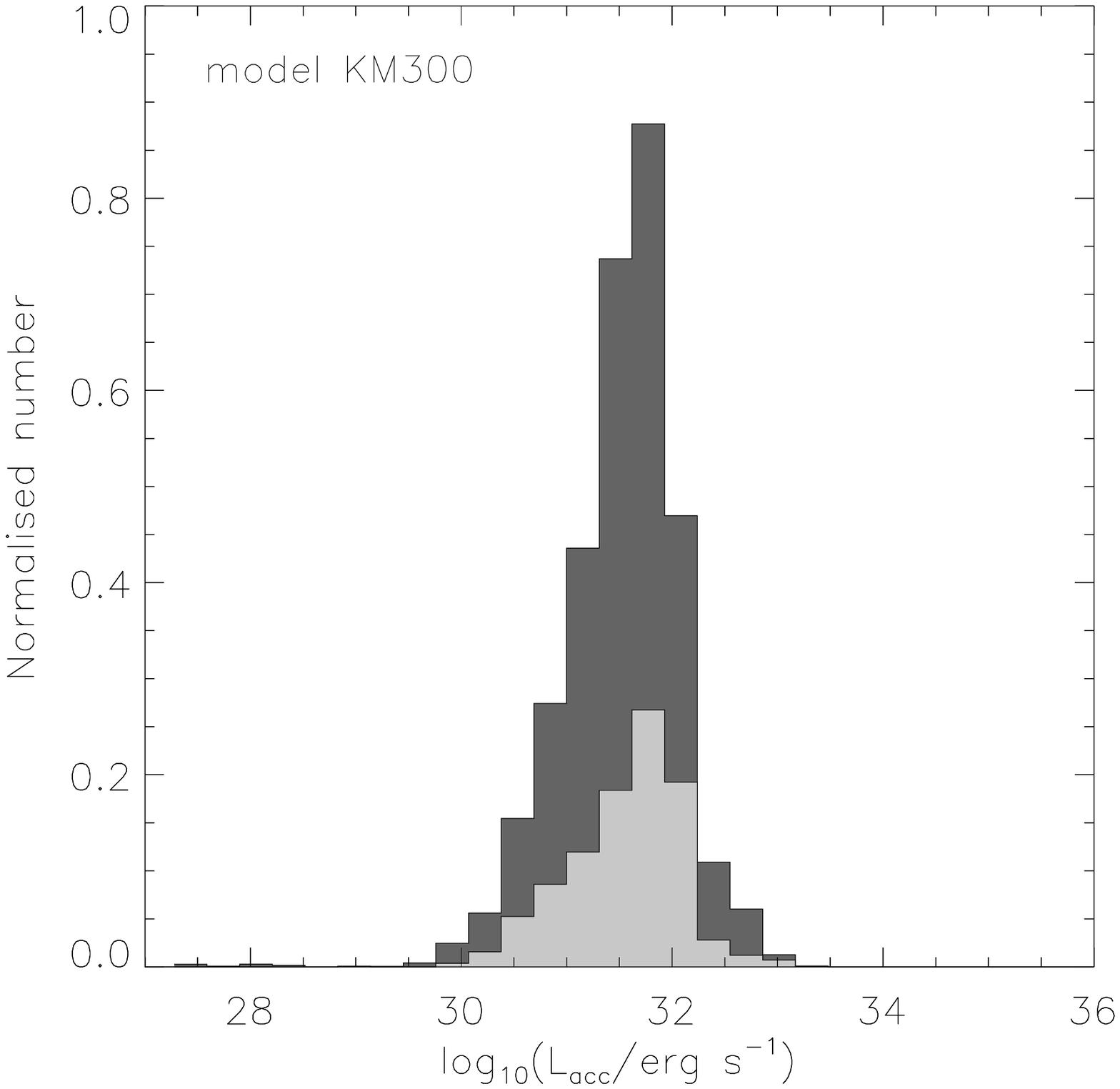}} \\
\resizebox{5.83cm}{!}{\includegraphics{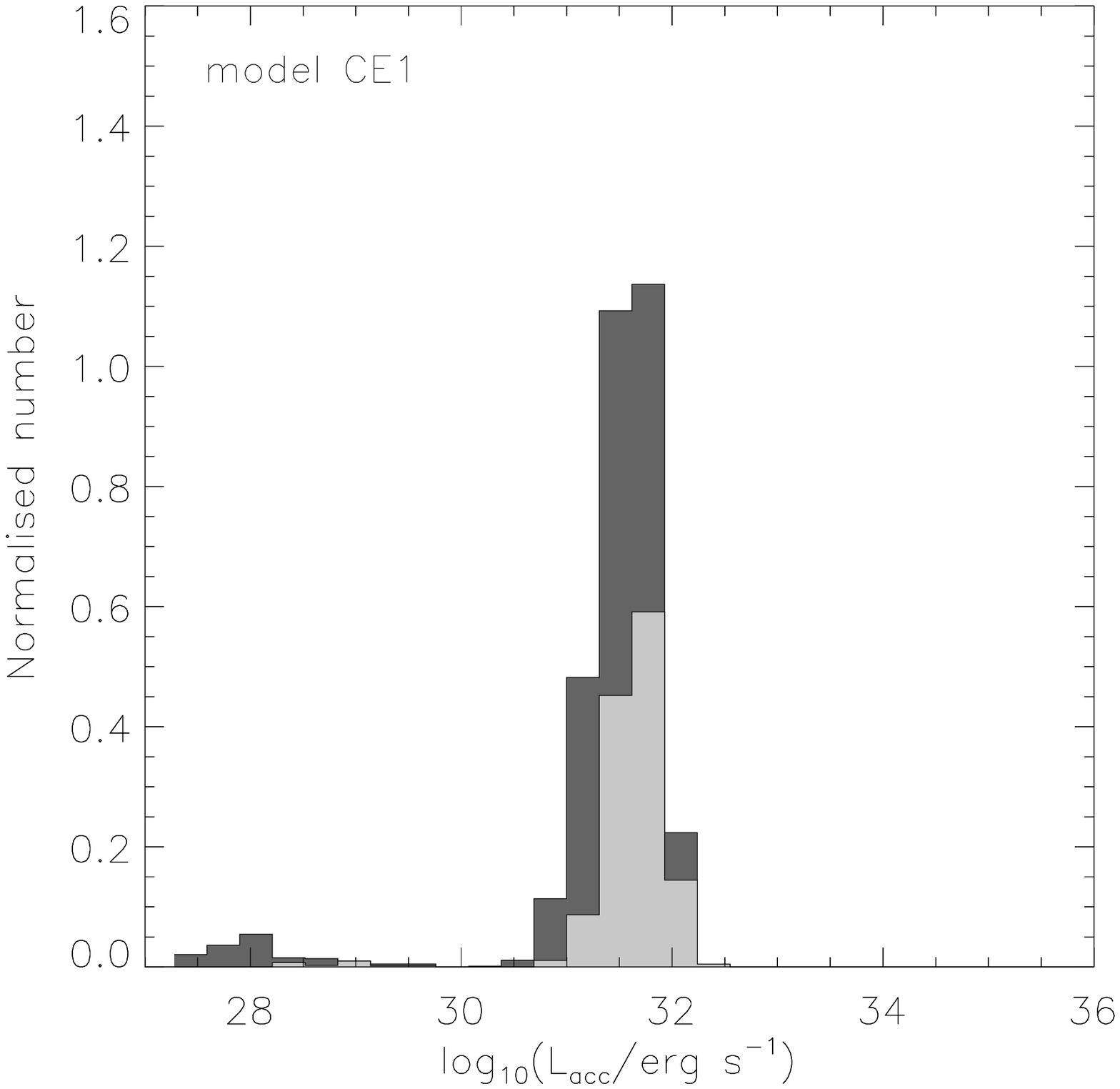}}
\resizebox{5.83cm}{!}{\includegraphics{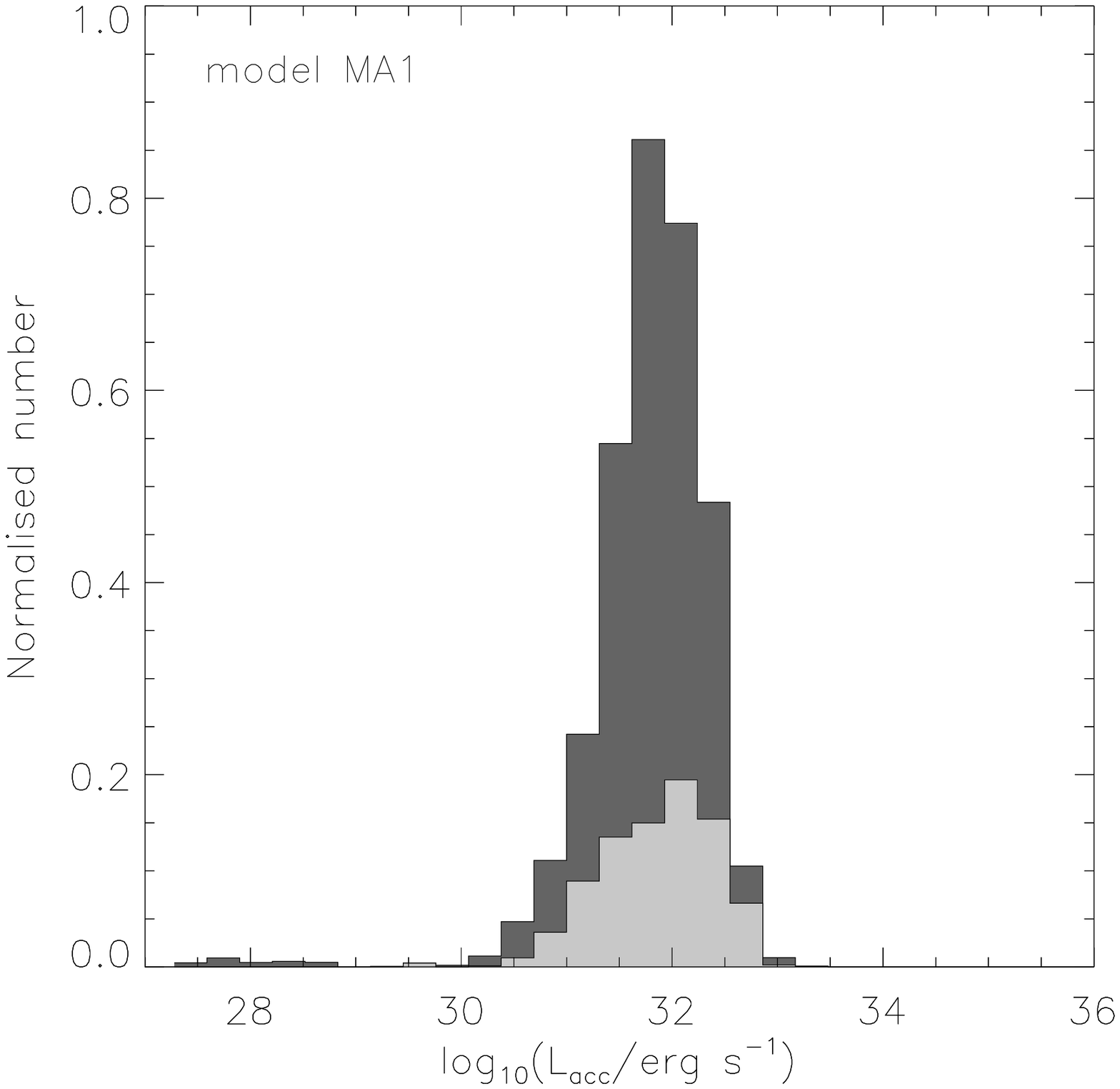}}
\resizebox{5.83cm}{!}{\includegraphics{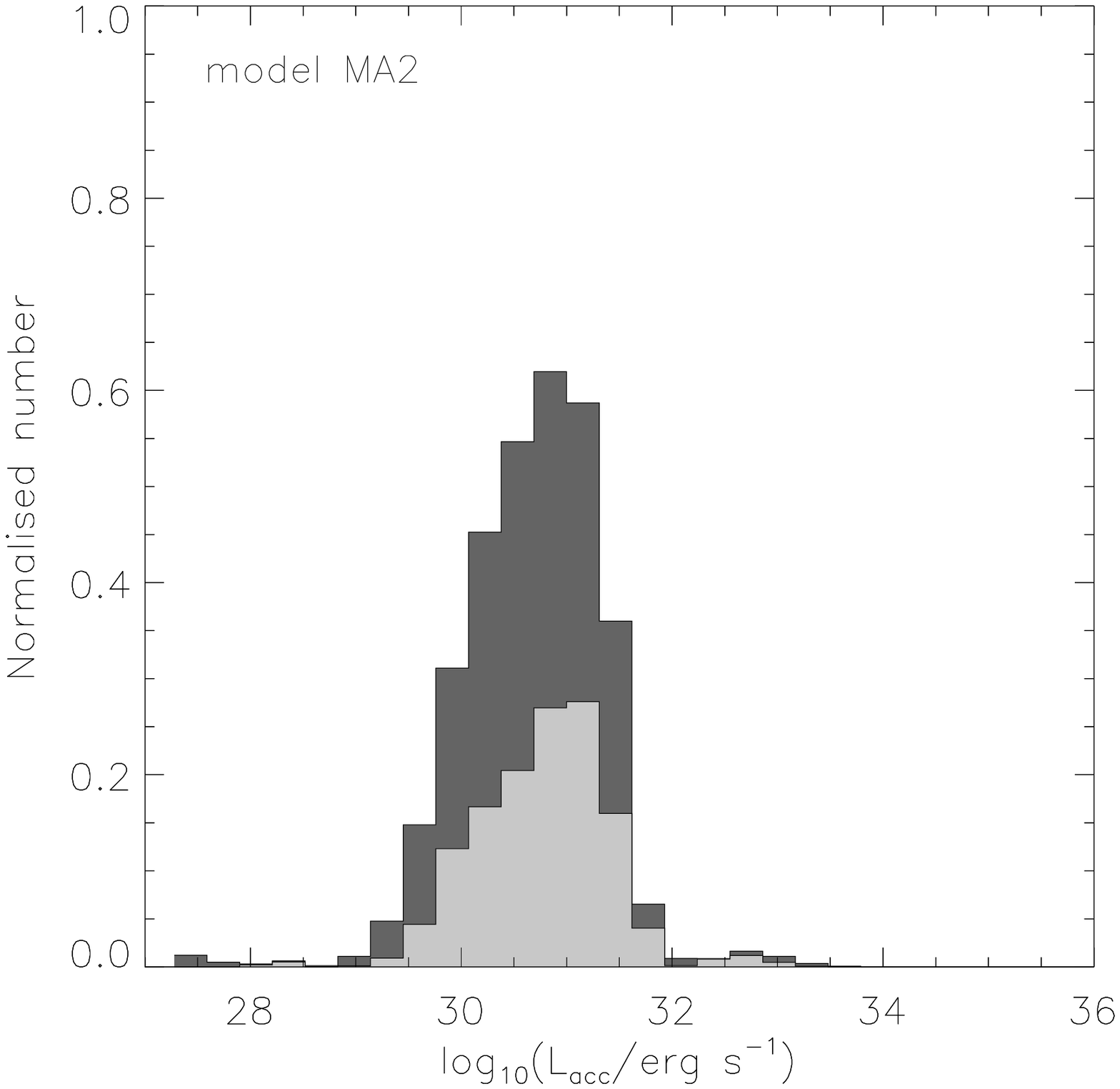}}
\caption{Normalised distribution functions of pre-LMXB accretion
  luminosities. Light and dark grey shadings correspond to the
  contributions of systems that do and do not evolve into LMXBs within
  the imposed age limit of 10\,Gyr, respectively. }
\label{L}
\end{figure*}

Finally, the one-dimensional distribution functions of the
pre-LMXB accretion luminosities are presented in Fig.~\ref{L} for the
same population synthesis models as in Fig.~\ref{MP}. We distinguish
between systems that actually evolve into LMXBs within the imposed age
limit of 10\,Gyr and those that do not by means of light and dark grey
shadings, respectively. The distribution of pre-LMXB accretion
luminosities in our standard population synthesis model shows a
primary peak around $10^{31}$ erg/s and a small secondary peak around
$10^{28}$ erg/s. The latter becomes more important with decreasing
kick-velocity dispersions and dominates when $\sigma_{\rm kick}=25$
km/s. This shift towards lower accretion luminosities is related to
the increasing contribution of the long-period systems with $P_{\rm
orb} \ga 1000$ days. The width of the peak at $10^{31}$ erg/s depends
primarily on the common-envelope ejection efficiency parameter
$\alpha_{\rm CE}$. The peak becomes narrower with decreasing values of
$\alpha_{\rm CE}$ due to the shrinking range of post-supernova orbital
periods around $P_{\rm orb} \approx 10$ days. Increasing the
wind-velocity parameter $\beta_{\rm w}$ moves the distribution
function to lower X-ray luminosities, while a decrease in $\beta_{\rm
w}$ has the opposite effect. A similar tendency was found by
Pfahl et al. (2002). It is also interesting to note that the peak 
luminosity of $10^{31}$ erg/s is comparable to the luminosity 
originating from non-magnetic white dwarfs accreting mass from a Roche-lobe 
filling companion (cataclysmic variables). Such systems may be 
distinguishable from pre-LMXBs based on the harder X-ray
spectrum expected for wind-accreting neutron stars. In addition, white
dwarfs can possibly be detected in the optical or UV wavelengths,
provided that they are bright enough. 

\section{Relative and absolute number of pre-LMXBs in the Galaxy}

The relative number of pre-LMXBs evolving into LMXBs within the
imposed age limit of 10\,Gyr are listed in the first column of
Table~\ref{frac}.  In general, about 20 to 30 per cent of all systems
satisfying our pre-LMXB criteria evolve into a LMXB in a time span of
less than 10\,Gyr. This fraction decreases significantly for
kick-velocity dispersions smaller than $\sim 50$ km/s (model KM50) due
to the growing contribution of systems with orbital periods longer
than $\sim 1000$ days. A decrease of the common-envelope efficiency
parameter $\alpha_{\rm CE}$ (model CE1) on the other hand increases
the relative number of pre-LMXBs evolving into LMXBs since it boosts
the contribution of the short-period systems. Similarly, increasing
the wind-velocity parameter $\beta_{\rm w}$ (model MA2) decreases the
accretion luminosities of the longer period systems below the $L_{\rm
acc} > 10\,L_{\rm cor}$ threshold so that again the relative number of
pre-LMXBs evolving into LMXBs increases. If Eq.~(\ref{Lcor}) is
replaced by a constant coronal X-ray luminosity of $10^{-2}\,L_\odot$
(model COR1), a large fraction of the low-mass systems with $M_{\rm d}
\la 1\,M_\odot$ can no longer satisfy the $L_{\rm acc} > 10\,L_{\rm
cor}$ criterion. The disappearance of these systems causes an increase
of the relative number of pre-LMXBs that will evolve into LMXBs.

\begin{table}
\caption{Relative number of pre-LMXBs evolving into
  LMXBs within the imposed age limit of 10\,Gyr and relative number of
  LMXBs descending from a pre-LMXB.}
\label{frac}
\begin{tabular}{lcc}
\hline
      & pre-LMXBs evolving & LMXBs descending \\
model & into LMXBs         & from pre-LMXBs \\
\hline
A     & 31.9\% & 43.5\% \\
K0    & 21.3\% &  0.1\% \\
KM25  & 12.7\% & 14.3\% \\
KM50  & 16.1\% & 10.8\% \\
KM75  & 24.1\% &  8.5\% \\
KM100 & 28.7\% &  9.7\% \\
KM300 & 30.1\% & 48.7\% \\
KM400 & 29.5\% & 48.2\% \\
CE1   & 40.7\% & 14.9\% \\
CE2   & 33.2\% & 73.6\% \\
MA1   & 26.1\% & 36.9\% \\
MA2   & 41.2\% & 39.1\% \\
COR1  & 45.8\% & 43.2\% \\
COR2  & 24.7\% & 43.5\% \\
COR3  & 26.5\% & 43.5\% \\
\hline
\end{tabular}
\end{table}

Although a detailed population synthesis study of the formation
and evolution of LMXBs is beyond the scope if this investigation, it
is interesting to consider the relative number of LMXBs that are
expected to evolve through a phase satisfying our pre-LMXB
criteria. The relative numbers are listed in the second column of
Table~\ref{frac} and typically range 30 to 50 per cent. The LMXBs that
did not evolve through a pre-LMXB phase all descend from intermediate
mass X-ray binaries (IMXB) with initial donor star masses outside the
mass range considered in this investigation (see also King \& Ritter
1999; Podsiadlowski \& Rappaport 2000; Kolb et al. 2000; Pfahl,
Rappaport \& Podsiadlowski 2003). A substantial fraction of LMXBs thus
evolves through an X-ray bright wind-accretion phase prior to
the onset of Roche-lobe overflow from the neutron star's companion.

For conclusion, we estimate the order of magnitude of the
birthrate and the absolute number of pre-LMXBs currently populating
the Galaxy. To this end, we assume all stars to be in binaries and we
normalise the star formation rate to produce a Galactic type II
supernova rate of $10^{-2}$ per year (Cappellaro, Evans \& Turatto
1999). The resulting birthrates and absolute numbers are listed in
Table~\ref{num}. Based on these calculations, we expect the current
population of pre-LMXBs in the Galactic disk to exist of about
$10^4$--$10^5$ systems, so that they may account for a non-negligible
part of the X-ray output in large scale surveys performed by Chandra
and XMM.

\begin{table}
\caption{Birthrates and total number of Galactic pre-LMXBs for
  each of the population synthesis models listed in
  Table~\ref{models}. The models are normalised to a Galactic type II
  supernova rate of $10^{-2}$ per year.}
\label{num}
\begin{tabular}{lcc}
\hline
model & birthrate (yr$^{-1}$) & total number \\
\hline
A     & $2.1 \times 10^{-6}$ & $5.0 \times 10^4$  \\
K0    & $1.6 \times 10^{-7}$ & $7.9 \times 10^4$  \\
KM25  & $3.9 \times 10^{-7}$ & $6.5 \times 10^4$  \\
KM50  & $2.7 \times 10^{-7}$ & $4.3 \times 10^4$  \\
KM75  & $2.3 \times 10^{-7}$ & $3.3 \times 10^4$  \\
KM100 & $4.3 \times 10^{-7}$ & $3.6 \times 10^4$  \\
KM300 & $1.9 \times 10^{-6}$ & $3.7 \times 10^4$  \\
KM400 & $1.2 \times 10^{-6}$ & $2.3 \times 10^4$  \\
CE1   & $2.4 \times 10^{-7}$ & $1.2 \times 10^4$  \\
CE2   & $7.5 \times 10^{-6}$ & $2.0 \times 10^5$  \\
MA1   & $3.3 \times 10^{-6}$ & $6.5 \times 10^4$  \\
MA2   & $8.5 \times 10^{-7}$ & $4.0 \times 10^4$  \\
COR1  & $3.2 \times 10^{-7}$ & $3.2 \times 10^4$  \\
COR2  & $3.8 \times 10^{-6}$ & $7.0 \times 10^4$  \\
COR3  & $3.3 \times 10^{-6}$ & $6.5 \times 10^4$  \\
\hline
\end{tabular}
\end{table}

\section{Discussion}

We have investigated the population of candidate pre-low-mass X-ray
binaries according to the detection mechanism proposed by Bleach
(2002). The mechanism is based on the search for low-mass stars with
X-ray luminosities that are too large to be attributed to the stellar
coronal activity, and on the subsequent identification of neutron star
binaries amongst these stars. The X-ray excess in these binaries 
can then possibly be ascribed to the accretion of mass by the neutron
star from the stellar wind of its companion.

In our study, we used the BiSEPS binary population synthesis code
described by Willems \& Kolb (2002). We limited ourselves to neutron
star binaries with main-sequence, Hertzsprung-gap, or giant-branch
donor stars of mass $M_{\rm d} \la 2\,M_\odot$ and simulated the X-ray
excess by the requirement that the X-ray luminosity generated by
the wind-accreting neutron star be at least 10 times larger than the
coronal X-ray luminosity of the mass-losing star. The X-ray
luminosity associated with the accretion process was determined under
the assumptions that no accretion-inhibiting effects such as strong
centrifugal forces are present and that all gravitational potential
energy of the infalling matter is converted into X-ray radiation. 
The maxiumum coronal X-ray luminosity for a star of a given mass and
radius was assumed to be proportional to the square of the star's
radius.

It follows that, in the case of low supernova kick-velocity
dispersions, the population of pre-LMXBs can be divided into two groups
of systems: a long-period group with $P_{\rm orb} \ga 1000$ days and a
short-period group with $P_{\rm orb} \approx 10$ days. The division
yields a bimodal distribution for the accretion luminosities of the
candidate pre-LMXBs. The long-period systems produce a peak in the
distribution around $10^{28}$ erg/s, while the short-period systems
produce a peak near $10^{31}$ erg/s.  For kick-velocity disperions
larger than $\sim 100$ km/s only the peak near $10^{31}$ erg/s
remains. The luminosity at which the distribution peaks furthermore
depends sensitively on the velocity of the donor star's wind.

These results are derived under the assumption that the stellar and
orbital parameters, and in particular the orbital period and the
accretion luminosity, of pre-LMXBs remain unchanged after their
formation. This assumption breaks down for binaries with $M_{\rm d}
\la 1\,M_\odot$ and $P_{\rm orb} \la 1$ day in which the evolution is
governed by the loss of angular momentum via gravitational radiation
and/or magnetic braking, and for binaries with $M_{\rm d} \ga
1\,M_\odot$ in which the evolution is governed by the nuclear
evolution of the donor star.

First, the decrease of the orbital period resulting from the
angular momentum loss in binaries with $M_{\rm d} \la 1\,M_\odot$ and
$P_{\rm orb} \la 1$ day yields an increase of the mean mass-accretion
rate and thus of the accretion luminosity generated by the neutron
star. However, from Eqs.~(\ref{Lacc}) and~(\ref{Macc2}) it follows
that an increase of the accretion luminosity by a factor of 10
requires a decrease of the orbital period by a factor of $\sim
5.6$. For a pre-LMXB with an orbital period of $\sim 1$ day, this
comes down to a decrease of the period to $\sim 4.3$ hours. Since
neutron star binaries with secondaries more massive than $\sim
0.5\,M_\odot$ becomes semi-detached before they can reach such short
periods, we can expect that the neglect of the orbital evolution does
not affect our estimates for the accretion luminosities by more than
an order of magnitude.

Second, the nuclear evolution of donor stars more massive than
$\sim 1\,M_\odot$ may yield a significant increase of the star's
radius and wind mass-loss rate within the age of the Galaxy. The
effect of both these evolutionary aspects is to increase the mean
mass-accretion rate of the neutron star. However, for main-sequence
stars less massive than $2\,M_\odot$, the increase of the stellar
radius increases the mean mass-accretion rate by less than an order of
magnitude [see Eqs.~(\ref{Lacc}) and (\ref{Macc2})]. The overall
effect on our results can therefore again be expected to be small. The
main uncertainty for main-sequence donor stars consequently stems from
their poorly understood wind mass-loss rates (e.g. Wargelin \& Drake
2001). Pre-LMXBs with donor stars that have evolved beyond the
main-sequence, on the other hand, have better understood mass-loss
rates, but contribute too little to the total population of newborn
pre-LMXBs to have a significant effect on the distribution functions
presented. In addition, systems with an orbital period of the order of
a few days will evolve into LMXBs shortly after the secondary leaves
the main sequence. A significant increase in the X-ray accretion
luminosity is therefore only possible for wider systems which do not
initiate Roche-lobe overflow until the secondary is well on its way
towards the tip of the giant branch or for systems which do not
initiate Roche-lobe overflow at all. For these systems, we may
underestimate the maximum accretion luminosity by several orders of
magnitude, but since the life time of a star on the giant branch is much
shorter than its main sequence life time, the contribution of the
systems to the pre-LMXB population during this phase is smaller than
their contribution when the donor star was still on the main
sequence. In addition, the wider orbital separations required for this
scenario somewhat counteract the increase in the accretion luminosity
associated with the evolution of the donor star.

Finally, we estimated the total number of pre-LMXBs currently
populating the Galaxy to be of the order of $10^4$--$10^5$. For
comparison, the estimated number of RS CVn type binaries in the Galaxy
is of the order of $10^6$ (Watson 1990) and the theoretically
predicted number of cataclysmic variables in the Galaxy is of the
order of $10^6-10^7$ (Kolb \& Willems 2003, in preparation). Both
these types of systems have X-ray luminosities comparable to the
$10^{31}$ erg/s peak luminosity found in our models for candidate
pre-LMXBs. In addition, the distribution of X-ray luminosities at
$10^{28}$ erg/s may be expected to be dominated by coronally active
stars. Our predicted number of pre-LMXBs satisfying the X-ray excess
criterion is therefore small in comparison to the number of these
other low-luminosity X-ray sources, but nevertheless high enough to
possibly account for an interesting fraction of the low-luminosity
discrete X-ray sources observed by Chandra and XMM. A systematic
search for candidate pre-LMXBs based on the X-ray excess criterion
can in principle be initiated by comparing exisiting X-ray (ROSAT,
Einstein, Chandra and XMM) and optical (Hipparcos, Tycho, HST GSC)
source catalogs. Simultaneous X-ray and optical observations by XMM
provide an attractive alternative.

\section*{Acknowledgements}
We thank Jarrod Hurley, Onno Pols, and Chris Tout for sharing
their SSE software package; and Philipp Podsiadlowski, Firoza
Sutaria and Robin Barnard for useful discussions. The referee, 
Peter Wheatley, is thanked for his constructive remarks. This
research was supported by the British Particle Physics and Astronomy
Research Council (PPARC).

\bsp

\label{lastpage}

\end{document}